x
\documentclass[12pt]{article}
\usepackage[T1]{fontenc}
\usepackage{geometry}
\usepackage{setspace}
\makeatletter

\providecommand{\LyX}{L\kern-.1667em\lower.25em\hbox{Y}\kern-.125emX\@}

 \newcommand{\lyxaddress}[1]{
   \par {\raggedright #1 
   \vspace{1.4em}
   \noindent\par}
 }

\makeatother
\def\slash#1{\setbox0=\hbox{$#1$}#1\hskip-\wd0\hbox to\wd0{\hss\sl/\/\hss}}
\begin{document}

\title{{\large NONLOCAL REGULARIZATION AND SPONTANEOUSLY BROKEN ABELIAN GAUGE THEORIES
FOR AN ARBITRARY GAUGE PARAMETER }\large }

\author{Anirban Basu }

\maketitle

\lyxaddress{{\large Department of Physics,University of Chicago,Chicago IL 60637 }\large }

\author{~~~~~~~~~~~~~~~~~~~~~~~~~~~~~~~~~~Satish
D. Joglekar }

\lyxaddress{{\large Department of Physics,Indian Institute of Technology,Kanpur 208016
{[}INDIA{]}}\large }

\begin{abstract}
{\large We study the non-local regularization for the case of a spontaneously
broken abelian gauge theory in the R\( _{\xi }- \)gauge with an arbitrary gauge
parameter \( \xi  \). We consider a simple abelian-Higgs model with chiral
couplings as an example. We show that if we apply the nonlocal regularization
procedure {[}to construct a nonlocal theory with FINITE mass parameter {]} to
the spontaneously broken R\( _{\xi }- \)gauge Lagrangian, using the quadratic
forms as appearing in this Lagrangian, we find that a physical observable in
this model, an analogue of the muon anomalous magnetic moment ,evaluated to
order O {[}g\( ^{2}] \) does indeed show \( \xi - \) dependence.We then apply
the modified form of nonlocal regularization that was recently advanced and
studied for the unbroken non-abelian gauge theories and discuss the resulting
WT identities and \( \xi - \)independence of the S-matrix elements.}{\large
\par} \end{abstract} \section{{\large INTRODUCTION}\large }

{\large The Standard Model {[}SM{]} is a renormalizable local quantum field
theory {[}QFT{]} with a local gauge invariance {[}1{]}. Calculations in the
SM require the use of a regularization that respects its local gauge invariance.There
are various schemes of regularization appropriate for nonabelian gauge theories:
the dimensional regularization {[}DR{]} ,and also a combination of Pauli-Villars
and higher derivatives were employed in the early gauge theory literature {[}2{]}.
However, most attempts to go beyond standard model involve supersymmetry for
which a regularization so efficient as the DR cannot be employed.New regularizations
have been proposed in the last decade that can be employed in supersymmetric
field theories and these are nonlocal regularization {[}3,4,5,6,7{]} and
differential regularization .For a comparative statement regarding the
benefits of the nonlocal regularization vis-a-vis other regularizations,the
reader is referred to the Introduction of Ref.{[}6{]} and to Ref.5 In this
work, we study nonlocal regularization further in some of its aspects.
}{\large \par}
{\large Nonlocal regularization proposed by Moffat {[}3{]} and Evans et al {[}4{]}
and given an elegant formulation by Kleppe and Woodard {[}5,6{]} has been studied
extensively since {[}7,8{]}. Renormalization procedure has been established
upto two loop order{[}5{]} in scalar theories.In Ref.{[}6{]},it has been shown
how the nonlocally regularized field theory can be constructed from a local
QFT in a systematic fashion. More importantly, it has been established there
that the local/global symmetries can be preserved in their nonlocal form and
the WT identities of local QFT's derivable from local symmetries such as gauge
invariance/BRS symmetry find their natural nonlocal extensions. This has been
done for the abelian gauge theories to all orders {[}2,6{]} and for nonabelian
gauge theories in Feynman gauge {[}6{]} upto one loop order {[}limited,in particular,
by the existence of measure beyond one loop{]}.The nonlocal theories have also
found applications in the clearer understanding of the renormalization program{[}14{]}.}{\large \par}

{\large As mentioned earlier, we shall study a further aspect of non-local regularization
which we shall now elaborate on.In the work of Ref. {[}6{]},a procedure for
nonlocalization of local Lagrangians was presented. It was further applied to
the nonlocalization of the nonabelian gauge theories and to the preservation
of its BRS symmetry in a nonlocal form for the Feynman gauge.It was
emphasized that the nonlocal theories with a finite \( \Lambda  \) can
themselves be looked upon as physical theories {[}3,6,11{]}. The significance
of this view-point has been elaborated in ref.9 to which the reader is
referred to for more details. We shall work in the context of these. In the
Ref. {[}9{]}, which we shall refer to as I henceforth,we studied nonlocal
regularization of pure nonabelian unbroken gauge theories with an arbitrary
gauge parameter \( \xi \) . For an arbitrary \( \xi  \), we established the
necessity of a modified treatment for constructing the nonlocal Lagrangian, as
compared to that given in {[}6{]} for the Feynman gauge. We ,then, gave a way
to regularize the nonabelian gauge theory for an arbitrary \( \xi  \) and
established the WT-identities that would imply the \( \xi  \)-independence of
the observables for this theory.Now, the SM calculations are calculations in
a} \emph{\large spontaneously broken} {\large gauge theory and that is what is
relevant in actual calculations of S-matrix elements. Also, in an unbroken
theory, the S-matrix elements may not exist due to infrared divergences and it
is harder to test the \( \xi  \)-independence there. In view of these facts
,therefore, we wish to study in this work, the nonlocal regularization in a}
\emph{\large spontaneously broken} {\large model and for} \emph{\large an
arbitrary} {\large \( \xi  \),. Apart from its relevance as mentioned above,
we find we have a few other motivations. In a spontaneously broken
abelian-Higgs model with chiral couplings to be introduced in Sec.3 {[} which
in many ways mimics the GWS model, yet is simpler to deal with{]}, we
explicitly perform, in this work,the calculation of a simple 1-loop
observable: the analogue of the g-2 for the muon in the model with an
arbitrary \( \xi  \) and find a \( \xi \)-dependent result. An analogous
calculation was done earlier by G. Saini and one of us {[}10{]} in the context
of an actual SM calculation {[}11{]} and we had found a \( \xi \)-dependent
result there. The above model of Sec.3 is studied here to explicitly bring out
this in a much simpler model with essentially those features that lead to this
\( \xi  \)-dependence: viz. spontaneously broken model with chiral couplings.
With this motivation in mind,we study the application of the modified way of
nonlocal regularization proposed and formulated in unbroken pure Yang-Mills
theories to the spontaneously broken theory in I, and study the WT-identities
in the theory and the \( \xi \)-independence of the resulting Green's
functions. }{\large \par} {\large We now present the plan of the paper. In
Section two, we present the earlier results, viz.:(a) the results in the Ref.
6 on the nonlocalization of a local theory that in particular,preserves the
local BRS invariance of the theory in the Feynman gauge,(b) some of the
essential results on the nonlocalization procedure, and (c) the conclusions of
I regarding the need to modify the general procedure outlined in {[}6{]} while
dealing with a general \( \xi  \) and also with calculations in the Feynman
gauge in higher loops. In section three, we introduce the abelian- Higgs model
with the chiral couplings for an arbitrary \( \xi  \) and introduce
nonlocalization of the R\( _{\xi } \)- gauge local Lagrangian following the
procedure of the Ref.6. In section four, we evaluate the analogue of the
1-loop contribution to the (g-2) of the muon in this model with a finite
mass scale and show that it exhibits \( \xi  \)-independence.In section 5 we
study the application of the modified regularization used in I and study the
WT-identity.We evaluate \( \frac{\partial W}{\partial \xi } \) and simplify
the result using the WT-identities. In section 6, we then discuss the \( \xi -
\)independence of the S-matrix elements.}{\large \par} \section{{\large
PRELIMINARIES}\large } {\large In this section, we shall introduce the
notations and review the earlier results on the non-local regularization of
reference 6 and of I. For brevity , we rely heavily on I , to which the reader
is often referred to for any further details. }{\large \par}
{\large (A) NON-LOCAL REGULARIZATION: }{\large \par}

{\large We shall, very briefly, review the essential properties of the non-local
regularization.We keep the introduction to bare essentials; referring the reader
to I for a more detailed introduction. }{\large \par}

{\large We write the local action for a field theory ,in terms of a generic
field \( \phi  \), as the sum of the quadratic and the interaction part: }{\large \par}

{\large S{[}\( \phi ] \)=F{[}\( \phi ] \)+I{[}\( \phi ] \)~~~~~~~~~~~~~~~~~~~~~~~~~~~~~~~~~~~~~~~~~~~~~~~~~~~~~~~~~~~~~~~~(2.1)}{\large \par}

{\large and the quadratic piece is expressed as }{\large \par}

{\large F{[}\( \phi ] \)= \( \int  \) d\( ^{4}x \) \( \phi _{i}(x)\Im _{ij}\phi _{j}(x) \)
~~~~~~~~~~~~~~~~~~~~~~~~~~~~~~~~~~~~~~~~~~~~~~~~~~~~~~~~~~~~~~~(2.2)}{\large \par}

{\large We define the regularized action in terms of the smeared field \( \widehat{\phi } \)
,defined in terms of the kinetic energy operator \( \Im _{ij} \) as,}{\large \par}

{\large \( \widehat{\phi } \) = \( \varepsilon  \)\( ^{-1}\phi  \) ~~~~~~~~~~~~~~~~~~~~~~~~~~~~~~~~~~~~~~~~~~~~~~~~~~~~~~~~~~~~(2.3)}{\large \par}

{\large with \( \varepsilon  \)= exp \{\( \Im  \) /\( \Lambda ^{2}\} \).The
nonlocally regularized action is constructed by first introducing an auxiliary
action \( S \){[}\( \phi ,\psi ] \).It is given by }{\large \par}

{\large \( S \){[}\( \phi ,\psi ] \)= F{[}\( \widehat{\phi } \){]} -A{[}\( \psi ] \)+
I {[}\( \phi + \)\( \psi ] \) ~~~~~~~~~~~~~~~~~~~~~~~~~~~~~~~~~~~~~~~~~~~~~~~~~~~~~~~~~~~~(2.4)}{\large \par}

{\large where \( \psi  \)is called a {}``shadow field{}'' with an action }{\large \par}

{\large A{[}\( \psi ] \)=\( \int  \) d\( ^{4}x \) \( \psi _{i}O^{-1}_{ij}\psi _{j} \)
~~~~~~~~~~~~~~~~~~~~~~~~~~~~~~~~~~~~~~~~~~~~~~~~~~~~~~~~~~~~(2.5)}{\large \par}

{\large with \( O \) defined by }{\large \par}

{\large \( O \) = {[}\( \frac{\varepsilon ^{2}-1}{\Im } \){]}~~~~~~~~~~~~~~~~~~~~~~~~~~~~~~~~~~~~~~~~~~~(2.5a)}{\large \par}

{\large The action of the non-local theory is defined as }{\large \par}

{\large \( \widehat{S} \) {[}\( \phi ] \)= \( S \){[}\( \phi ,\psi ] \)\( \Vert _{\psi =\psi [\phi ]} \)
~~~~~~~~~~~~~~~~~~~~~~~~~~~~~~~~~~~~~~~~~~~~~~~~~~~~~~~~~~(2.6)}{\large \par}

{\large where \( \psi [\phi  \){]} if the solution of the classical equation}{\large \par}

{\large 0= \( \frac{\delta S}{\delta \psi } \)~~~~~~~~~~~~~~~~~~~~~~~~~~~~~~~~~~~~~~~~~~~~~~~~~~~~~~~~~~(2.7)}{\large \par}

{\large The nonlocalized Feynman rules are simple extensions of the local ones.
The vertices are unchanged but every leg can connect either to a smeared propagator}{\large \par}

{\large \( \frac{i\varepsilon ^{2}}{\Im +i\epsilon } \)=-i\( \int  \)\( _{_{1}} \)\( ^{^{\infty }} \)\( \frac{d\tau }{\Lambda ^{2}}exp\{\frac{\Im }{\tau \Lambda ^{2}}\} \)~~~~~~~~~~~~~~~~~~~~~~~~~~~~~~~~~~~~~~~~~~~~~~~~~~~~~~~~~(2.8)}{\large \par}

{\large or to a shadow propagator {[}a line crossed by a bar{]}}{\large \par}

{\large \( \frac{i[1-\varepsilon ^{2}]}{\Im +i\epsilon }= \)-iO =-i \( \int  \)\( ^{1} \)\( _{_{0}}\frac{d\tau }{\Lambda ^{2}}exp\{\frac{\Im }{\tau \Lambda ^{2}}\} \)~~~~~~~~~~~~~~~~~~~~~~~~~~~~~~~~~~~~~~~~~~~~~~(2.9)}{\large \par}

{\large A set of detailed comments on the Feynman rules and the diagrams that
need to be considered are found in I.We only need to remember that in a Feynman
diagram, the internal lines can be either shadow or smeared lines with the exception
that no diagrams with closed shadow loops are counted.}{\large \par}

{\large A number of theorems that finally relate to the construction of the
nonlocal symmetry transformations of the nonlocal theory given the local ones
for the local theory are established in {[}6{]} and are found enumerated in
I. We only recall that related to the nonlocal BRS symmetry.}{\large \par}

{\large Theorem B.3:{[}We stick to the theorem numbering of {[}6{]} {]}. If
S\{{[}\( \phi  \){]}\} is invariant under \( \delta \phi _{i} \)= T\( _{i} \)
{[}\( \phi ] \), then \( \widehat{S} \) {[}\( \phi ] \) is invariant under}{\large \par}

{\large \( \widehat{\delta } \)\( \phi  \)\( _{i} \)= \( \varepsilon ^{2}_{ij} \)
T\( _{j} \) {[}\( \phi  \)+\( \psi  \){[}\( \phi  \){]}{]}~~~~~~~~~~~~~~~~~~~~~~~~~~~~~~~~~~~~~~~~~~~~~~~~~~~~~~~~(2.10)}{\large \par}

{\large (B) NONLOCAL REGULARIZATION OF A GAUGE THEORY FOR AN ARBITRARY GAUGE
PARAMETER}{\large \par}

{\large We shall first review briefly the results, in Ref. 6, on nonlocalization
of the nonabelian gauge theories in the Feynman gauge .We consider the Feynman
gauge local effective action :}{\large \par}

{\large S\( _{F}= \)\( \int  \) d\( ^{4}x \) {[}-\( \frac{1}{2} \)\( \partial _{\mu }A_{\nu } \)
\( \partial ^{\mu }A^{\nu } \)-\( \partial ^{\mu }\overline{\eta } \)\( \partial _{\mu }\eta  \){]}
+ I {[}\( A^{\mu },\overline{\eta } \)\( ,\eta  \){]}~~~~~~~~~~~~~~~~~~~~~~~~~~~~~~~~~~~~~~~~~~~~~~~~~~~~~~~~(2.11)}{\large \par}

{\large Ref. 6 outlines the results as to how a local action is converted to
a non-local action.Briefly put,it is done in the present context by first noting
the quadratic operators \( \Im _{ij} \) = \( \partial  \)\( ^{2} \)\( \delta  \)\( ^{4} \)(x-y)\( \eta  \)\( _{\mu \nu } \)
for the gauge field and \( \partial  \)\( ^{2} \)\( \delta  \)\( ^{4} \)(x-y)
for the ghost fields and employing these to construct the smeared fields \( \widehat{A_{\mu }} \)
=exp\{\( -\partial  \)\( ^{2} \)/\( \Lambda ^{2}\} \) A\( _{\mu } \) ;\( \widehat{\eta } \)=exp\{\( -\partial  \)\( ^{2} \)/\( \Lambda ^{2}\} \)\( \eta  \);\( \widehat{\overline{\eta }} \)=exp\{\( -\partial  \)\( ^{2} \)/\( \Lambda ^{2}\} \)\( \overline{\eta } \).
In terms of these,and the corresponding shadow fields B\( _{\mu } \)\( ^{a} \),\( \psi  \)\( ^{a} \),
\( \overline{\psi ^{a}} \);we have the non-local action:}{\large \par}

{\large \( \widehat{S_{F}} \)=\( \int  \) d\( ^{4}x \) {[}-\( \frac{1}{2} \)\( \partial _{\mu }\widehat{A_{\nu }} \)\( \partial ^{\mu }\widehat{A^{\nu }} \)-\( \partial ^{\mu }\widehat{\overline{\eta }} \)\( \partial _{\mu }\widehat{\eta } \)
-\( \frac{1}{2} \)B\( ^{a} \)\( _{\mu } \)O\( ^{-1} \)B\( ^{a} \)\( ^{\mu } \)+
\( \overline{\psi ^{a}} \)O\( ^{-1} \)\( \psi  \)\( ^{a} \){]} + I {[}\( A^{\mu }+B^{\mu },\overline{\eta } \)+
\( \overline{\psi } \)\( ,\eta  \)+\( \psi  \){]} ~~~~~~~~~~~~~~~~~~~~~~~~~~~~~~~~~~~~~~~~~~~~~~~~~~~~~~~(2.12)}{\large \par}

{\large The local action of (2.11) has a local BRS symmetry.The non-local action
,correspondingly, has a} \emph{\large nonlocal} {\large BRS symmetry obtained
from the result in Sec.2(A). As emphasized in Ref.6, it is found convenient
to combine this with a {}``trivial{}'' i.e. equation of motion symmetry so
that the resultant BRS variation of \( \overline{\eta } \) is proportional
to \( \partial .A^{a} \),the field corresponding to the longitudinal degree
of freedom.This helps in establishing the WT that establishes the decoupling
of the longitudinal degree of freedom.These nonlocal BRS transformations are:}{\large \par}

{\large \( \widehat{\delta } \)A\( ^{a} \)\( _{\mu } \)= {[} \( \partial _{\mu } \)\( \eta  \)\( ^{a} \)
-g f\( ^{abc} \) \( \varepsilon ^{2} \) (A\( _{\mu } \) + B\( _{\mu } \)
)\( ^{b} \)\( (\eta  \)+\( \psi ) \)\( ^{c} \){]}\( \delta \zeta  \)~~~~~~~~~~~~~~~~~~~~~~~~~~~~~~~~~~~~~~~~~~~~~~~~(2.13a)}{\large \par}

{\large \( \widehat{\delta } \)\( \eta ^{a} \) =- \( \varepsilon ^{2} \)\( \frac{g}{2} \)f\( ^{abc} \)\( \eta  \)\( ^{b}\eta  \)\( ^{c} \)\( \delta \zeta  \)~~~~~~~~~~~~~~~~~~~~~~~~~~~~~~~~~~~~~~~~~~~~~~~~~~~~~~(2.13b)}{\large \par}

{\large \( \widehat{\delta } \)\( \overline{\eta } \)\( ^{a} \)=-\( \partial .A^{a} \)\( \delta \zeta  \)
~~~~~~~~~~~~~~~~~~~~~~~~~~~~~~~~~~~~~~~~~~~~~~~~~~~~~(2.13c)}{\large \par}

{\large As emphasized in I, however, when this procedure of nonlocalization
as outlined in Sec 2(A) is applied in the straightforward manner, to the case
of Yang-Mills theory with an arbitrary gauge parameter \( \xi  \) , it fails
to yield WT identities that would yield a \( \xi  \)-independent observables.Moreover,the
procedure outlined above in Ref. {[}6{]} cannot be employed directly to go beyond
1-loop even for the Feynman gauge,as the gauge parameter \( \xi  \) gets renormalized
and in effect the action in terms of the bare parameters does contain a gauge
parameter \( \neq  \)1. As a consequence, a formulation valid for all \( \xi  \),
must be employed if one is to go beyond 1-loop in this case also.In I, we presented
a formulation valid for all \( \xi  \) and established the relevant WT identities
that would imply \( \xi  \)-independence of observables.To illustrate this
consider the Yang-Mills effective action for an arbitrary \( \xi  \):}{\large \par}

{\large \( \widehat{S} \)\( _{\xi } \)=\( \widehat{S} \)\( _{F} \)+\( \Delta  \)\( \widehat{S} \) }{\large \par}

{\large where \( \widehat{S} \)\( _{F} \) has been given in (2.12) and }{\large \par}

{\large \( \Delta  \)\( \widehat{S} \) = (1-1/\( \xi  \))\( \int  \) d\( ^{4}x \)
(\( \partial  \).\( \widehat{A} \) )\( ^{2} \) }{\large \par}

{\large The smeared field operators depend only on the quadratic form in \( \widehat{S} \)\( _{F} \).
Hence, for an arbitrary \( \xi  \), we have \( \widehat{A} \)\( _{\mu } \)=
(\( \varepsilon ^{-1}_{F} \))\( _{\mu \nu } \)A\( ^{\nu } \)=exp\{\( -\partial  \)\( ^{2} \)/\( \Lambda ^{2}\} \)
A\( _{\mu } \) and \( \widehat{\eta } \)=(\( \varepsilon ^{-1}_{\eta } \))\( \eta  \)=e
xp\{\( -\partial  \)\( ^{2} \)/\( \Lambda ^{2}\} \)\( \eta  \);\( \widehat{\overline{\eta }} \)=exp\{\( -\partial  \)\( ^{2} \)/\( \Lambda ^{2}\} \)\( \overline{\eta } \).
We note that \( \varepsilon _{F}= \)\( \varepsilon _{\eta }\equiv  \)\( \varepsilon  \)
here. Further,\( O \) = {[}\( \frac{\varepsilon ^{2}-1}{\Im } \){]} is the
same for B and for \( \psi  \) and is independent of \( \xi  \). The nonlocal
BRS symmetry for the case of an arbitrary \( \xi  \) has been obtained in I.It
reads , for an arbitrary \( \xi  \),}{\large \par}

{\large \( \widehat{\delta } \)A\( ^{a} \)\( _{\mu } \)= {[} \( \partial _{\mu } \)\( \eta  \)\( ^{a} \)
-g f\( ^{abc} \) \( \varepsilon ^{2} \) (A\( _{\mu } \) + B\( _{\mu } \)
)\( ^{b} \)\( (\eta  \)+\( \psi ) \)\( ^{c} \){]}\( \delta \zeta  \)~~~~~~~~~~~~~~~~~~~~~~~~~~~~~~~~~~~~~~~~~~~~~~~~(2.14a)}{\large \par}

{\large \( \widehat{\delta } \)\( \eta  \)\( ^{a} \) =- \( \varepsilon ^{2} \)\( \frac{g}{2} \)f\( ^{abc} \)\( \eta  \)\( ^{b}\eta  \)\( ^{c} \)\( \delta \zeta  \)~~~~~~~~~~~~~~~~~~~~~~~~~~~~~~~~~~~~~~~~~~~~~~~~~~~~~~(2.14b)}{\large \par}

{\large \( \widehat{\delta } \)\( \overline{\eta } \)\( ^{a} \)=-\( \frac{1}{\xi } \)\( \partial .A^{a} \)\( \delta \zeta  \)
~~~~~~~~~~~~~~~~~~~~~~~~~~~~~~~~~~~~~~~~~~~~~~~~~~~~~(2.14c)}{\large \par}

{\large In I,we have discussed the WT identities arising from this symmetry
and have seen that it leads to a result that would imply the \( \xi  \)-independence
of those S-matrix related physical quantities that exist.}{\large \par}

\section{{\large NONLOCAL REGULARIZATION OF A SPONTANEOUSLY BROKEN ABELIAN MODEL WITH
CHIRAL FERMIONS}\large }

{\large In this work,we shall study an aspect of non-local regularization as
related to the spontaneously broken gauge theories. In order to do so,we shall
first introduce in this section a simple model - an abelian Higgs Model with
Chiral couplings.It is perhaps worthwhile to elaborate and to iterate the context
in which we found it essential to study this model.In the work of Ref. {[}6{]},a
procedure for nonlocalization of nonabelian gauge theories and preservation
of the BRS symmetry in a nonlocal form was presented for the Feynman gauge.
Subsequently, in I, we studied nonlocal regularization of pure nonabelian unbroken
gauge theories with an arbitrary gauge parameter \( \xi  \) . For an arbitrary
\( \xi  \), we established the necessity of a modified treatment for constructing
the nonlocal Lagrangian as compared to that given in {[}6{]} for the Feynman
gauge. We ,then, gave a way of regularize the nonabelian gauge theory for an
arbitrary \( \xi  \) and established the WT-identities that would imply the
\( \xi  \)-independence of the observables for this theory. In view of the
fact that the calculation of an observable in the SM generally involves a spontaneously
broken gauge theory ,the extension of an analogous treatment to the case of
spontaneously broken case is of value.In this work,therefore, we wish to study
, for an arbitrary \( \xi  \), the nonlocal regularization in a spontaneously
broken model. Apart from its relevance as mentioned above, we find we have a
few other motivations. In section IV, we shall perform the calculation of a
simple 1-loop observable: the analogue of the g-2 for the muon in the model
with arbitrary \( \xi  \) explicitly and find a \( \xi  \)-dependent result.
An analogous calculation was done earlier by G. Saini and one of us {[}10 {]}
in the context of an actual SM calculation{[}11{]} and we had found a \( \xi  \)-dependent
result.The above model is studied here to explicitly bring out this in a simpler
model with essentially those features that lead to this: viz. spontaneously
broken model with chiral couplings. With this motivation in mind,we study the
application of the modified way of nonlocal regularization proposed and formulated
in unbroken pure Yang-Mills theories to the spontaneously broken theory in I
and study the WT-identities in the theory and the \( \xi  \)-independence of
the resulting Green's functions. Therefore, in this section,we shall introduce
an abelian Higgs model with gauge fields coupling to vector and the axial vector
currents.As mentioned earlier, this simpler model incorporates in it several
essential features of the standard model such as coupling of gauge fields to
chiral fermions and a spontaneously broken nature of the model. Both these features
are essential in the example that we shall discuss in the next section.}{\large \par}

{\large Consider the Lagrangian for the model with a local U(1)XU(1) symmetry
,having a complex scalar \( \varphi  \) ,a fermion \( \psi  \) and two U(1)
gauge fields A and B , coupling respectively to vector and the axial vector
currents and the quantum numbers for the fields as indicated below:}\footnote{%
The model below has the chiral anomaly.It can be removed by adding another fermion
\( \Psi  \) with electric charge \( \pm 1 \) and opposite axial charge. Introduction
of \( \Psi  \) does not alter the one-loop calculation of (g-2) for \( \psi  \)
presented in the next section. Hence, to this order we ignore the question of
the anomaly .
}{\large \par}

\vspace{0.375cm}
{\centering \begin{tabular}{|c|c|c|}
\hline 
{\large }&
{\large U\( _{V}(1) \)}&
{\large U\( _{A} \)\( (1) \)}\\
\hline 
\hline 
{\large \( \varphi  \)}&
{\large 0}&
{\large 2}\\
\hline 
{\large \( \psi  \)\( _{L} \)}&
{\large 1}&
{\large 1}\\
\hline 
{\large \( \psi  \)\( _{R} \)}&
{\large 1}&
{\large -1}\\
\hline 
\end{tabular}\large \par}
\vspace{0.375cm}

{\large ~~~~~~~~~~~~~~~~~~~~~~~~~~~~~~~~~~~~~Table
1:Quantum Numbers}{\large \par}

{\large L\( _{inv} \)= (D\( _{\mu }\varphi  \))\( ^{\dagger } \)(D\( ^{\mu }\varphi  \))-
\( \mu ^{2}\varphi ^{\dagger }\varphi  \)-\( \lambda  \)\( (\varphi ^{\dagger }\varphi )^{2} \)-1/4
F\( _{\mu \nu } \)F\( ^{\mu \nu } \)-1/4 G\( _{\mu \nu } \)G\( ^{\mu \nu } \)+
\( \overline{\psi }_{L}(i{\slash{\partial}}  \)  +g{\slash{B}}-e{\slash{A}})\( \psi  \)\( _{L} \)+\( \overline{\psi }_{R}(i{\slash{\partial}}  \)-g{\slash{B}}-e{\slash{A}})\( \psi  \)\( _{R} \) + f\( (\overline{\psi }_{L} \)\( \psi  \)\( _{R} \)\( \varphi  \)+\( \overline{\psi _{R}} \)\( \psi  \)\( _{L} \)\( \varphi  \){*})~
~~ ~~ ~~ ~~ ~~ ~~ ~~ ~~ ~~ ~~ ~~ ~~ ~~ ~~ ~~ ~(3.1)}{\large \par}

{\large where,}{\large \par}

{\large \( \varphi = \)\( \frac{\varphi _{1}+i\varphi _{2}}{\sqrt{2}} \) ;
\( \varphi  \)\( _{1} \), and \( \varphi  \)\( _{2} \) real.}{\large \par}

{\large D\( _{\mu } \) =\( \partial _{\mu } \)-2igB\( _{\mu } \) }{\large \par}

{\large F\( _{\mu \upsilon } \)=\( \partial _{\mu } \)A\( _{\nu } \)\( -\partial _{\nu }A_{\mu } \)}{\large \par}

{\large G\( _{\mu \upsilon } \)=\( \partial _{\mu } \)B\( _{\nu } \)\( -\partial _{\nu }B_{\mu } \)
;~ ~~ ~~ ~~ ~~ ~~ ~~ ~~ ~~ ~~ ~~ ~~ ~~ ~~ ~~ ~~
(3.2)}{\large \par}

{\large the (mass)\( ^{2} \)parameter\( =\mu  \)\( ^{2} \) < 0 and \( \varphi  \)
develops a vacuum expectation value given by:}{\large \par}

{\large <\( \varphi ^{\dagger }\varphi  \)>= -\( \mu  \)\( ^{2} \)/2\( \lambda  \)\( \equiv  \)\( \frac{v^{2}}{2} \)
> 0 .~ ~~ ~~ ~~ ~~ ~~ ~~ ~~ ~~ ~~ ~~ ~~ ~~ ~~ ~~
~~(3.3)}{\large \par}

{\large Further, the gauge-fixing Lagrangian in the R\( _{\xi } \)-gauges is
given by}\footnote{%
We keep the gauge parameter associated the unbroken U(1) field equal to 1.
}{\large \par}

{\large L\( _{gf} \)=-1/2 (\( \partial .A) \)\( ^{2} \) -\( \frac{1}{2\xi } \)(\( \partial  \).B
+ \( \xi  \)M\( \varphi  \)\( _{2} \))\( ^{2} \), M\( \equiv  \)2gv~~
~~ ~~ ~~ ~~ ~~ ~~ ~~ ~~ ~~ ~~ ~~ ~~ ~~ ~~(3.4)}{\large \par}

{\large Thus,we can now re-express the net Lagrangian in terms of the shifted
fields: \( \varphi  \)\( _{1}' \)=\( \varphi  \)\( _{1} \)-v, and \( \varphi  \)\( _{2} \)'=\( \varphi  \)\( _{2} \)
as L=L\( _{0} \)+L\( _{I} \) ,with:}{\large \par}

{\large L\( _{0} \)= \( \frac{1}{2} \){[}(\( \partial _{\mu }\varphi _{1}' \))\( ^{2} \)
+ 2\( \mu  \)\( ^{2} \)\( \varphi _{1} \)'\( ^{2} \){]} \( +\frac{1}{2} \){[}(\( \partial _{\mu }\varphi _{2}' \))\( ^{2} \)-M\( ^{2} \)\( \xi \varphi _{2} \)'\( ^{2} \){]}
-\( \frac{1}{4} \)(\( \partial _{\mu } \)B\( _{\nu }- \)\( \partial _{\mu } \)B\( _{\nu } \))\( (\partial ^{\mu } \)B\( _{^{^{^{\nu }}}} \)-\( \partial ^{\nu } \)B\( ^{\mu } \))
+ \( \frac{1}{2} \)M\( ^{2} \)B\( _{\mu } \)B\( ^{\mu } \)-\( \frac{1}{2\xi } \)(\( \partial  \).B)\( ^{2} \)-\( \frac{1}{2} \)\( \partial _{\mu } \)A\( _{\nu } \)\( \partial ^{\mu } \)A\( _{^{^{^{\nu }}}} \)+\( \overline{\psi }(i{\slash{\partial}}  \)-m)\( \psi  \)
~~ ~~ ~~ ~~ ~~ ~~ ~~ ~~ ~~ ~~ ~~ ~~ ~~ ~~ ~(3.5a)}{\large \par}

{\large L\( _{I} \) = \( \overline{\psi }_{L}( \)g{\slash{B}}-e{\slash{A}})\( \psi  \)\( _{L} \)+\( \overline{\psi }_{R}( \)-g{\slash{B}}-e{\slash{A}})\( \psi  \)\( _{R} \)-\( \frac{m}{v} \)\( \overline{\psi }_{L} \)\( \psi  \)\( _{R} \)(\( \varphi _{1}' \)+i\( \varphi  \)\( _{2}') \)-\( \frac{m}{v} \)\( \overline{\psi _{R}} \)
\( \psi  \)\( _{L} \)(\( \varphi _{1}' \)-i\( \varphi  \)\( _{2}') \) }{\large \par}

{\large +2gB\( ^{\mu } \) (\( \varphi _{2}'\partial _{\mu }\varphi _{1}'- \)\( \varphi _{1}'\partial _{\mu }\varphi _{2}') \)
+ 2 g\( ^{2} \)B\( _{\mu } \)B\( ^{\mu } \) ( \( \varphi _{1}'^{2} \)+\( \varphi _{2}'^{2}) \)
-\( \frac{\lambda }{4} \)( \( \varphi _{1}'^{2} \)+\( \varphi _{2}'^{2}) \)\( ^{2} \)
+ 4 g\( ^{2} \)vB\( _{\mu } \)B\( ^{\mu } \)\( \varphi  \)\( _{1}' \)-\( \lambda  \)v\( \varphi _{1}' \)(
\( \varphi _{1}'^{2} \)+\( \varphi _{2}'^{2}) \) ;~~ ~~ ~~ ~~ ~~
~~ ~~ ~~ ~~ ~~ ~~ ~~ ~~ ~~ ~(3.5b)}{\large \par}

{\large here m=-fv/\( \sqrt{2} \) stands for the fermion mass.}{\large \par}

{\large We now assume}\footnote{%
In the case of spontaneously broken gauge theories,there are potentially two
ways of nonlocalising. The one is as shown here in this work. The other is to
start from the model in an unbroken form, nonlocalize the model and then break
symmetry.In the latter case,we cannot ,of course, add the gauge fixing in the
R\( _{\xi } \)-form in the unbroken action. However, the entire procedure in
the latter case turns out to be very cumbersome.
} {\large that the procedure for nonlocalizing the above local Lagrangian is
as elaborated in Sec. 2(A).To this end, we recall the quadratic structures in
L\( _{0} \) and introduce the shadow fields for each of the fields in L as
indicated in the following table:}{\large \par}

\vspace{0.375cm}
{\centering \begin{tabular}{|c|c|c|c|c|c|}
\hline 
{\large fields}&
{\large \( \varphi  \)\( _{1}' \)}&
{\large \( \varphi  \)\( _{2}' \)}&
{\large \( \psi  \)}&
{\large A}&
{\large B}\\
\hline 
\hline 
{\large \( shadow \) fields}&
{\large \( \rho _{1} \)}&
{\large \( \rho _{2} \)}&
{\large \( \varsigma  \)}&
{\large H}&
{\large D}\\
\hline 
\end{tabular}\large \par}
\vspace{0.375cm}

{\large ~~~~~~~~~~~~~~~~~~~~~~~~~~~~~~~~~~~~~~~~~~~~~~
Table 2}{\large \par}

\vspace{0.375cm}
{\centering \begin{tabular}{|c|c|}
\hline 
{\large field}&
{\large propagator}\\
\hline 
\hline 
{\large \( \varphi _{1}' \)}&
\multicolumn{1}{|c|}{{\large -i\( \int  \)\( ^{\infty }_{_{1}} \)\( \frac{d\tau }{\Lambda ^{2}} \)exp
\{ \( \frac{\tau }{\Lambda ^{2}} \)(k\( ^{2} \)+2\( \mu ^{2} \))\}}}\\
\hline 
{\large \( \varphi _{2}' \)}&
{\large -i\( \int  \)\( ^{\infty }_{_{1}} \)\( \frac{d\tau }{\Lambda ^{2}} \)exp
\{ \( \frac{\tau }{\Lambda ^{2}} \)(k\( ^{2} \)-M\( ^{2}\xi  \))\}}\\
\hline 
{\large \( \psi  \)}&
{\large -i\( \int  \)\( ^{\infty }_{_{1}} \)\( \frac{d\tau }{\Lambda ^{2}} \)exp
\{ \( \frac{\tau }{\Lambda ^{2}} \)(k\( ^{2} \)-m\( ^{2} \))\} ({\slash{k}} +m)}\\
\hline 
{\large A}&
{\large i\( \int  \)\( ^{\infty }_{_{1}} \)\( \frac{d\tau }{\Lambda ^{2}} \)exp
\{ \( \frac{\tau }{\Lambda ^{2}} \)k\( ^{2} \)\} \( \eta _{\mu \nu } \)}\\
\hline 
{\large B}&
{\large i\( \int  \)\( ^{\infty }_{_{1}} \)\( \frac{d\tau }{\Lambda ^{2}} \)exp
\{ \( \frac{\tau }{\Lambda ^{2}} \)(k\( ^{2} \)-\( M^{2} \))\}{[}\( \eta _{\mu \nu } \)-\( \frac{k_{\mu }k_{\nu }}{k^{2}}\{1-exp( \)-\( \frac{\tau }{\Lambda ^{2}} \)k\( ^{2} \)(1-\( \frac{1}{\xi }))\} \){]};}\\
\hline 
\end{tabular}\large \par}
\vspace{0.375cm}

{\large ~~~~~~~~~~~~~~~~~~~~~~~~~~~~~~~~~~~~~~~~~~~~~~~
Table 3}{\large \par}

\vspace{0.375cm}
{\centering \begin{tabular}{|c|c|}
\hline 
{\large \( shadow \) field}&
{\large propagator}\\
\hline 
\hline 
{\large \( \rho _{1} \)}&
{\large -i\( \int  \)\( ^{^{1}} \)\( _{_{0}} \)\( \frac{d\tau }{\Lambda ^{2}} \)exp
\{ \( \frac{\tau }{\Lambda ^{2}} \)(k\( ^{2} \)+2\( \mu ^{2} \))\};}\\
\hline  
{\large \( \rho _{2} \)}&
{\large -i\( \int  \)\( ^{^{1}}_{_{0}} \)\( \frac{d\tau }{\Lambda ^{2}} \)exp
\{ \( \frac{\tau }{\Lambda ^{2}} \)(k\( ^{2} \)-M\( ^{2}\xi  \))\};}\\
\hline 
{\large \( \varsigma  \)}&
{\large -i\( \int  \)\( ^{^{1}}_{_{0}} \)\( \frac{d\tau }{\Lambda ^{2}} \)exp
\{ \( \frac{\tau }{\Lambda ^{2}} \)(k\( ^{2} \)-m\( ^{2} \))\} ({\slash{k}} +m)}\\
\hline 
{\large H}&
{\large i\( \int  \)\( ^{^{1}} \)\( _{_{0}} \)\( \frac{d\tau }{\Lambda ^{2}} \)exp
\{ \( \frac{\tau }{\Lambda ^{2}} \)k\( ^{2} \)\}\( \eta _{\mu \nu } \);}\\
\hline 
{\large D}&
{\large i\( \int  \)\( ^{^{1}} \)\( _{_{0}} \)\( \frac{d\tau }{\Lambda ^{2}} \)exp
\{ \( \frac{\tau }{\Lambda ^{2}} \)(k\( ^{2} \)-\( M^{2} \))\}{[}\( \eta _{\mu \nu } \)-\( \frac{k_{\mu }k_{\nu }}{k^{2}}\{1-exp( \)-\( \frac{\tau }{\Lambda ^{2}} \)k\( ^{2} \)(1-\( \frac{1}{\xi }))\} \){]}}\\
\hline 
\end{tabular}\large \par}
\vspace{0.375cm}

{\large ~~~~~~~~~~~~~~~~~~~~~~~~~~~~~~~~~~~~~~~~~~~~~
Table 4}{\large \par}

\section{{\large A CALCULATION TO TEST \protect\( \xi \protect \)-INDEPENDENCE}\large }

{\large In this section,we shall present a calculation to test the \( \xi  \)-dependence
of a physical quantity.The quantity we consider is the analogue, in the context
of the present model,of the (g-2) of a muon in the standard model.The calculation
proceeds very much along the lines of the reference {[}11{]} ,where the (g-2)
of the muon was calculated in the non-local Standard model.The essential points
to note in this 1-loop calculation are : }{\large \par}

{\large (a) The quantity we are calculating is finite both in the local and
the nonlocal theories;hence so is the difference. This difference is of order
O{[}\( \frac{m^{2}}{\Lambda ^{2}} \){]} as compared to the local result for
(g-2).}{\large \par}

{\large (b) The difference between the local and the nonlocal results is given
by the various 1-loop diagrams having a shadow loop; and the total of these
as seen above are finite in value;}{\large \par}

{\large (c) As a result, the conclusion of the calculation we are to perform
below, is independent of a renormalization procedure;}{\large \par}

{\large (d) The local contribution is known to be \( \xi  \)-independent; so
that the \( \xi  \)-independence, or the lack of it, is determined by the}
\emph{\large finite 1-loop shadow diagrams.}{\large \par}

{\large We enumerate these diagrams involved .These are shown in Fig.1 and correspond
to the exchanges of H,D , \( \rho _{1,}\rho _{2} \).Noting that vertices are
\( \xi  \)-independent, a glance at the Table 4 in Section 3 will show that
the \( \xi  \)-dependence of the result can possibly come only from the D and
the \( \rho _{2} \)-exchange.These are the diagrams of Fig.1(b) and 1(d).}{\large \par}

{\large In evaluating the \( \xi  \)-dependence to the leading order, the following
observation about a loop integral comes handy: Consider an integral of the form }{\large \par}

{\large \( \frac{1}{\Lambda ^{4}} \)\( \int  \)\( \frac{d^{4}l}{(2\pi )^{4}} \)exp
\{\( \alpha  \)( \( \xi  \))\( l^{2} \)/\( \Lambda  \)\( ^{2} \) + \( \beta  \)(\( \xi  \))
m\( ^{2} \)/\( \Lambda  \)\( ^{2} \)\} ~~~~~~~~~~~~~~~~~~~~~~~~~~~~~~~~~~~~~~~~~~~~~~~~~~~~~~~~~~~~~(4.1)}{\large \par}

{\large We can evaluate the above integral by expanding exp\( \{\beta  \)(\( \xi  \))
m\( ^{2} \)/\( \Lambda  \)\( ^{2} \)\} in powers of \( \beta  \)(\( \xi  \))
m\( ^{2} \)/\( \Lambda  \)\( ^{2} \) since each term is a finite integral.Then,
we have}{\large \par}

{\large \( \frac{1}{\Lambda ^{4}} \)\( \int  \)\( \frac{d^{4}l}{(2\pi )^{4}} \)exp
\{\( \alpha  \)( \( \xi  \))\( l^{2} \)/\( \Lambda  \)\( ^{2} \) + \( \beta  \)(\( \xi  \))
m\( ^{2} \)/\( \Lambda  \)\( ^{2} \)\} = \( \frac{A}{\alpha (\xi )^{2}} \)+
terms of higher order in 1/\( \Lambda  \)\( ^{2} \).~~~~~~~(4.2)}{\large \par}

{\large Thus, as far as the leading terms arising from (4.1) are concerned,
the contribution to the \( \xi  \)-dependence from \( \beta  \)(\( \xi  \))
m\( ^{2} \)/\( \Lambda  \)\( ^{2} \) can be ignored; and to the leading order,
the \( \xi  \)-dependence of (4.1 ) is determined by \( \alpha  \)( \( \xi  \)). }{\large \par}

{\large With this in mind, we turn our attention to the diagrams 1(b) and 1(d).We
recall the propagator of \( \rho _{2} \) in 1(d) from the Table 4. We shall
first show that the diagram of Fig.1(d) does not have \( \xi  \)-dependent
term on the leading order {[} O (m\( ^{2} \)/\( \Lambda  \)\( ^{2} \)) {]}.To
see this, we write out the expression for the diagram using the propagators
from Table 4: }{\large \par}

{\large i\( \Delta \Gamma  \)\( ^{(1)} \)\( _{\alpha } \)= \( \int  \)\( \frac{d^{4}l}{(2\pi )^{4}} \){[}-i\( \int  \)\( ^{^{1}}_{_{0}} \)\( \frac{d\tau _{3}}{\Lambda ^{2}} \)exp
\{ \( \frac{\tau _{3}}{\Lambda ^{2}} \)((p-l)\( ^{2} \)-M\( ^{2}\xi  \))\}{]}\( \overline{u(q)} \)\( \frac{m}{v}\gamma _{5} \)}{\large \par}

{\large \( \bullet [- \)i\( \int  \)\( ^{^{1}}_{_{0}} \)\( \int  \)\( \frac{d\tau _{1}}{\Lambda ^{2}} \)exp
\{ \( \frac{\tau _{1}}{\Lambda ^{2}} \)((\( l-p+q) \)\( ^{2} \)-m\( ^{2} \))\}(\( {\slash{l}}-{\slash{p}}+{\slash{q}} \)+m){]}(-i\( e\gamma _{\alpha } \)){[}-i\( \int  \)\( ^{^{1}}_{_{0}} \)\( \frac{d\tau _{2}}{\Lambda ^{2}} \)exp
\{ \( \frac{\tau _{2}}{\Lambda ^{2}} \)(\( l \)\( ^{2} \)-m\( ^{2} \))\}
(\( {\slash{l}} \)+m){]}\( \frac{m}{v}\gamma _{5} \)u(p)~~~~~~~~~~~~~~~~~~~~~~~~~~~~~~~~~~~~~~~~~~~~~~~~~~~~~~~~~~~~~~~~~~~~~~~~~~~~~~~~~~~~~~~~~~~~(4.3)}{\large \par}

{\large We now let \( l \)-->\( l \) + \( \frac{\tau _{1+}\tau _{3}}{\tau } \)\( p \)-
\( \frac{\tau _{1}}{\tau } \)\( q \); with \( \tau =\tau _{1+}\tau _{2+}\tau _{3} \).This
gives}{\large \par}

{\large i\( \Delta \Gamma  \)\( _{_{\alpha }}^{(1)} \)(p,q) =e\( [\frac{m}{v}] \)\( ^{2}\int  \)\( \frac{d^{4}l}{(2\pi )^{4}} \)\( \int  \)\( ^{^{1}}_{_{0}} \)\( \frac{d\tau _{1}}{\Lambda ^{2}} \)\( \int  \)\( ^{^{1}}_{_{0}} \)\( \frac{d\tau _{2}}{\Lambda ^{2}} \)\( \int  \)\( ^{^{1}}_{_{0}} \)\( \frac{d\tau _{3}}{\Lambda ^{2}} \)\( \overline{u(q)} \)\( \gamma _{5} \)(\( {\slash{l}} \)
-\( \frac{\tau _{2}}{\tau } \){\slash{p}} +\( \frac{\tau _{3+}\tau _{2}}{\tau } \){\slash{q}}+m)
\( \gamma _{\alpha } \)(\( {\slash{l}} \) +\( \frac{\tau _{1+}\tau _{2}}{\tau } \)\( {\slash{p}} \)-\( \frac{\tau _{1}}{\tau } \){\slash{q}}+m)\( \gamma _{5} \)u(p)
exp \{ \( \frac{\tau }{\Lambda ^{2}}\textrm{l} \)\( ^{2} \)-2\( \frac{\tau _{1}\tau _{2}}{\tau \Lambda ^{2}} \)\( p.q \)-\( \frac{1}{\Lambda ^{2}}[\tau _{3} \)M\( ^{2}\xi  \)+\( \frac{\tau _{1^{^{2}}+}\tau _{2}^{2}}{\tau }m^{2} \))\}
~~~~~~~~~~~~~~~~~~~~~~~~~~~~~~~~~~~~~~(4.4)}{\large \par}

{\large We now recall the remark made below (4.2). The above expression, as
result of this remark ,does not have a \( \xi  \)-dependent contribution to
(g-2) to the leading order {[} O (m\( ^{2} \)/\( \Lambda  \)\( ^{2} \)) {]}.}{\large \par}

{\large We now turn our attention to the diagram 1(b).The expression for this
shadow diagram is }{\large \par}

{\large \( \Delta \Gamma  \)\( _{\alpha }^{(2)} \)=\( \int  \)\( \frac{d^{4}l}{(2\pi )^{4}} \)\( \overline{u(q)} \)(-ig\( \gamma ^{\mu } \)\( \gamma _{5} \)){[}i\( \int  \)\( ^{^{1}} \)\( _{_{0}} \)\( \frac{d\tau _{1}}{\Lambda ^{2}} \)exp
\{ \( \frac{\tau _{1}}{\Lambda ^{2}} \)(\( l \)\( ^{2} \)-\( M^{2} \))\}{[}\( \eta _{\mu \nu } \)-\( \frac{l_{\mu }l_{\nu }}{l^{2}}\{1-exp( \)-\( \frac{\tau _{1}}{\Lambda ^{2}} \)\( l \)\( ^{2} \)(1-\( \frac{1}{\xi }))\} \){]}}{\large \par}

{\large \( \bullet [ \)-i\( \int  \)\( ^{^{1}} \)\( _{_{0}} \)\( \frac{d\tau _{2}}{\Lambda ^{2}} \)exp
\{ \( \frac{\tau _{2}}{\Lambda ^{2}} \)((\( l-q) \)\( ^{2} \)-m\( ^{2} \))\}(-\( {\slash{l}}+{\slash{q}} \)+m){]}(-i\( e\gamma _{\alpha } \)){[}-i\( \int  \)\( ^{^{1}} \)\( _{_{0}} \)\( \frac{d\tau _{3}}{\Lambda ^{2}} \)exp
\{ \( \frac{\tau _{3}}{\Lambda ^{2}} \)((p-\( l \))\( ^{2} \)-m\( ^{2} \))\}
({\slash{p}}-\({\slash{l}} \)+m){]}(-ig\( \gamma ^{\nu } \)\( \gamma _{5} \))u(p)~~~~~~~~~~~~~~~~~~~~~~~~~~~~~~~~~~~~~~~~~~~~~~~~~~~~~~~~~~~~~~~~~~~~~~~~~~~~~~~~~~~~~~~~~~(4.5)}{\large \par}

{\large For reasons mentioned in paragraph below eq.(4.2) , it is easily seen
that the \( \eta _{\mu \nu } \) term in the D-propagator cannot contribute
to the \( \xi  \)-dependence of the (g-2) in the leading order.Thus we need
to consider the contribution of the \( l_{\mu }l \)\( _{\nu } \)/\( l^{2} \)
term only. These terms contribute essentially due to the chiral nature of vertices
and give a contribution proportional to m\( ^{2} \).The result for terms that
contribute to (g-2) reads:}{\large \par}

{\large \( \Delta \Gamma  \)\( _{\mu }^{(2)} \) = \( \frac{e}{2m} \)\( \frac{g^{2}}{2\pi ^{2}} \)\( \frac{m^{2}}{\Lambda ^{2}} \)\( \overline{u(q)} \)
\( \sigma  \)\( _{\mu \nu } \)u(p) (p-q)\( ^{\nu } \)\( \int  \)\( ^{^{1}} \)\( _{_{0}} \)d\( \tau  \)\( _{1} \)\( \int  \)\( ^{^{1}} \)\( _{_{0}} \)\( d\tau _{2}\int  \)\( ^{^{1}} \)\( _{_{0}} \)d\( \tau  \)\( _{3} \) }{\large \par}

{\large \( \bullet  \){[} \( \xi  \)-independent terms from the \( \eta _{\mu \nu } \)-
term + \( \frac{\tau _{2}}{3[\tau +\tau _{1}(1-1/\xi )]^{3}} \)-\( \frac{\tau _{3}\tau _{2}}{4[\tau +\tau _{1}(1-1/\xi )]^{4}} \){]}+{[}
O (m\( ^{2} \)/\( \Lambda  \)\( ^{2} \))\( ^{2} \) {]}.}{\large \par}

{\large A direct evaluation of the above integral shows that it is \( \xi  \)-dependent.}{\large \par}

{\large We thus note the presence of \( \xi  \)-dependent terms of O{[}\( \frac{m^{2}}{\Lambda ^{2}} \){]}
which are the only \( \xi  \)-dependent terms contributing to (g-2).}{\large \par}

{\large We summarize the result of this section. We considered an abelian Higgs
model with axial and vector couplings. We then wrote down the spontaneously
broken R\( _{\xi } \)-gauge local action and nonlocalized the action using
the recipe of the reference 6 for an arbitrary \( \xi  \).We then considered
a physically measurable quantity viz. (g-2) of the fermion to check the \( \xi  \)-dependence
of the result.We note that the calculation of this quantity in 1-loop does not
depend on renormalizations that need to be carried out.We noted that the nonlocal
contribution was finite and given by the negative of the relevant shadow diagrams.
We isolated the diagrams that could be \( \xi  \)-dependent.We then showed
that the \( \rho _{2} \)-exchange diagram does not lead to a \( \xi  \)-dependence
in the leading order contribution to (g-2). We evaluated the contribution to
(g-2) from the diagram with the D-exchange and it turned out to have \( \xi  \)-dependence
in order O{[}\( \frac{m^{2}}{\Lambda ^{2}} \){]}.}{\large \par}

{\large Keeping in mind that the mass here has arisen due to spontaneous symmetry
breakdown, we see that the above demonstration of gauge-dependence has been
dependent on the two features of a spontaneously broken model and of having
the axial couplings:} \emph{\large both of which are present in the standard
model}{\large . In fact a calculation along the lines of reference 11, gives
a \( \xi  \)-dependent result for (g-2) of the muon in the standard model {[}10{]}
if the above procedure is used for nonlocalization there.}{\large \par}

\section{{\large NONLOCAL REGULARIZATION FOR AN ARBITRARY \protect\( \xi \protect \)
AND THE WT-IDENTITIES }\large }

{\large As shown in the previous section, the general procedure of nonlocalization,when
applied to the spontaneously broken local action in R\( _{\xi } \)-gauges for
the model considered there, leads to a \( \xi  \)-dependent physical observable
. In I , we had formulated an alternate procedure for nonlocalization of the
nonabelian gauge theories and established a WT-identity that would imply the
\( \xi  \)-independence of the physical observables. We wish to carry out an
analogous exercise here for the model presented in section 3. {[}We note that
since this is a spontaneously broken model,a larger class of S-matrix elements
themselves (rather than cross-sections with an energy cut-off) exist.{]}We shall
write down the WT identities for the model and obtain \( \frac{\partial W}{\partial \xi } \)\( \mid _{_{\xi =1}} \).Then,
in the next section,we shall discuss the gauge-independence of the S-matrix
using these results.}{\large \par}

{\large L\( _{BRS} \) =L\( _{inv} \)+L\( _{gf} \) +L\( _{gh} \) }{\large \par}

{\large =(D\( _{\mu }\varphi  \))\( ^{\dagger } \)(D\( ^{\mu }\varphi  \))-
\( \mu ^{2}\varphi ^{\dagger }\varphi  \)-\( \lambda  \)\( (\varphi ^{\dagger }\varphi )^{2} \)-\( \frac{1}{2} \)\( \partial _{\mu } \)A\( _{\nu } \)\( \partial ^{\mu } \)A\( _{^{^{^{\nu }}}} \)-1/4
G\( _{\mu \nu } \)G\( ^{\mu \nu } \)+ \( \overline{\psi }_{L}(i{\slash{\partial}}  \) +g{\slash{B}}-e{\slash{A}})\( \psi  \)\( _{L} \)}{\large \par}

{\large +\( \overline{\psi }_{R}(i{\slash{\partial}}  \)-g{\slash{B}}-e{\slash{A}})\( \psi  \)\( _{R} \)
+ \( \frac{\sqrt{2}m}{v} \)\( (\overline{\psi }_{L} \)\( \psi  \)\( _{R} \)\( \varphi  \)+\( \overline{\psi _{R}} \)\( \psi  \)\( _{L} \)\( \varphi  \){*})~
-\( \frac{1}{2} \)(\( \partial  \).B + \( \xi  \)M\( \varphi  \)\( _{2} \)')\( ^{2} \)+
\( \partial  \)\( ^{\mu } \)\( \overline{\eta } \) \( \partial _{\mu }\eta  \)-2Mg
\( \varphi _{1} \)\( \overline{\eta } \)\( \eta  \) ~~~~~~~~~~~~~~~~~~~~~~~~~~~~(5.1)}{\large \par}

{\large has the following BRS symmetry {[}arising from the U\( _{A}(1) \) local
symmetry{]} in the Feynman gauge \( \xi  \)=1:}{\large \par}

{\large \( \delta \varphi _{1} \)=2\( \eta  \)\( \varphi _{_{2}} \)\( \delta \zeta  \);
\( \delta \eta  \)=0 ; \( \delta B_{\mu } \)=-\( \frac{1}{g} \)\( \partial _{\mu }\eta  \)\( \delta \zeta  \) }{\large \par}

{\large \( \delta \varphi _{_{2}} \)=-2\( \eta  \)\( \varphi _{1} \)\( \delta \zeta  \)
; \( \delta \overline{\eta } \) =- \( \frac{1}{g} \)(\( \partial  \).B +
M\( \varphi  \)\( _{2} \))\( \delta \zeta  \); \( \delta A_{\mu } \)=0 }{\large \par}

{\large \( \delta \psi  \)= -i\( \eta \gamma _{5} \)\( \psi  \)\( \delta \zeta
\)~~~~~~~~~~~~~~~~~~~~~~~~~~~(5.2)}{\large \par}
{\large The nonlocal Feynman gauge action is obtained by using the quadratic
forms in (5.1) and following the procedure as in Sec 2. It reads, }{\large \par}

{\large \( \widehat{S} \)\( _{F} \) =\( \int  \)d\( ^{4}x \)\{\( \frac{1}{2}\widehat{\varphi _{1}'} \)(-\( \partial ^{2}+2\mu ^{2}) \)\( \widehat{\varphi _{1}'} \)+\( \frac{1}{2}\widehat{\varphi _{2}'} \)(-\( \partial ^{2}-M^{2}) \)
\( \widehat{\varphi _{2}'} \)+\( \frac{1}{2} \)\( \widehat{B} \)\( _{\mu } \)(\( \partial ^{2}+M^{2}) \)\( \widehat{B} \)\( ^{\mu } \)+
\( \frac{1}{2} \) \( \widehat{A} \)\( _{\mu } \) \( \partial  \)\( ^{2} \)\( \widehat{A} \)\( ^{\mu } \) }{\large \par}

{\large +\( \widehat{\overline{\psi }} \)i\( {\slash{\partial}}  \)\( \widehat{\psi } \)-m\( \widehat{\overline{\psi }} \)\( \widehat{\psi } \)-
\( \widehat{\overline{\eta }} \)\( \partial ^{2} \)\( \widehat{\eta } \)-2Mgv\( \widehat{\overline{\eta }} \)\( \widehat{\eta } \)\}-\( \int  \)d\( ^{4}x \)\{\( \frac{1}{2} \)\( \rho  \)\( _{1} \)O\( ^{-1} \)\( _{\varphi _{1}'} \)\( \rho  \)\( _{1} \)+\( \frac{1}{2} \)\( \rho  \)\( _{2} \)O\( ^{-1} \)\( _{\varphi _{2}'} \)\( \rho  \)\( _{2} \) }{\large \par}

{\large +\( \frac{1}{2} \)D\( ^{\mu } \)O\( ^{-1} \)\( _{B} \)D\( _{\mu } \)+\( \frac{1}{2} \)H\( ^{\mu } \)O\( ^{-1} \)\( _{A} \)H\( _{\mu } \)+\( \overline{\zeta } \)O\( ^{-1} \)\( _{\psi } \)\( \zeta  \)-\( \overline{c} \)O\( ^{-1} \)\( _{\eta } \)c\}
+ I ,~~~~~~~~~~~~~~~~~~~~~~~~~~~(5.3)}{\large \par}

{\large where the interaction term I is }{\large \par}

{\large I=\( \int  \)d\( ^{4}x \) \{-2Mg\( (\varphi _{_{1}} \)'+\( \rho  \)\( _{1} \))(\( \overline{\eta }+\overline{c} \)
)(\( \eta +c \) \( ) \) + 2g(B +D)\( ^{\mu } \)\( [(\varphi _{_{2}} \)'+\( \rho  \)\( _{2} \))\( \partial _{\mu } \)\( (\varphi _{_{1}} \)'+\( \rho  \)\( _{1} \))-\( (\varphi _{_{1}} \)'+\( \rho  \)\( _{1} \))\( \partial _{\mu } \)\( (\varphi _{_{2}} \)'+\( \rho  \)\( _{2} \)){]}}{\large \par}

{\large +2g\( ^{2} \)(B +D)\( ^{2} \){[}\( (\varphi _{_{2}} \)'+\( \rho  \)\( _{2} \))\( ^{2} \)\( + \)\( (\varphi _{_{1}} \)'+\( \rho  \)\( _{1} \))\( ^{2} \){]}-\( \frac{\lambda }{4} \){[}\( (\varphi _{_{2}} \)'+\( \rho  \)\( _{2} \))\( ^{2} \)\( + \)\( (\varphi _{_{1}} \)'+\( \rho  \)\( _{1} \))\( ^{2} \){]}\( ^{2} \)}{\large \par}

{\large +4g\( ^{2} \)v(B +D)\( ^{2} \)\( (\varphi _{_{1}} \)'+\( \rho  \)\( _{1} \))-\( \lambda  \)v\( (\varphi _{_{1}} \)'+\( \rho  \)\( _{1} \)){[}\( (\varphi _{_{2}} \)'+\( \rho  \)\( _{2} \))\( ^{2} \)\( + \)\( (\varphi _{_{1}} \)'+\( \rho  \)\( _{1} \))\( ^{2} \){]}}{\large \par}

{\large +(\( \overline{\psi }_{L}+\overline{\zeta _{L}})[ \)g({\slash{B}} +{\slash{D}})-e({\slash{A}} +{\slash{H}}){]} \( (\psi  \)\( _{L} \)\( +\zeta  \)\( _{L}) \)+(\( \overline{\psi }_{R}+\overline{\zeta _{R}})[ \)-g({\slash{B}}+ {\slash{D}})-e({\slash{A}} +{\slash{H}}){]}\( (\psi  \)\( _{R} \)\( +\zeta  \)\( _{R} \))
-\( \frac{m}{v} \)(\( \overline{\psi }_{L}+\overline{\zeta _{L}})[ \)\( \varphi _{_{1}} \)'+\( \rho  \)\( _{1} \)\( +i\varphi _{_{2}} \)'+i\( \rho  \)\( _{2} \){]}\( (\psi  \)\( _{R} \)\( +\zeta  \)\( _{R} \))}{\large \par}

{\large -\( \frac{m}{v} \)(\( \overline{\psi }_{R}+\overline{\zeta _{R}}) \)\( [ \)\( \varphi _{_{1}} \)'+\( \rho  \)\( _{1} \)\( -i\varphi _{_{2}} \)'-i\( \rho  \)\( _{2} \){]}\( (\psi  \)\( _{L} \)\( +\zeta  \)\( _{L}) \)~~~~~~~~~~~~~~~~~~~~~~~~~~~(5.4)}{\large \par}

{\large The nonlocalized symmetries of the nonlocal action in the Feynman gauge
then are}{\large \par}

{\large \( \widehat{\delta }\varphi _{1}' \)=2\( \varepsilon ^{2}_{\varphi _{1}'} \)\( (\varphi _{_{2}} \)'+\( \rho  \)\( _{2} \))(\( \eta +c \)
\( )\delta \zeta  \);}{\large \par}

{\large \( \widehat{\delta }\varphi _{2}' \)=-2\( \varepsilon ^{2}_{\varphi _{2}'} \)\( (\varphi _{_{1}} \)'+\( \rho  \)\( _{1} \)+v)(\( \eta +c \)
\( )\delta \zeta  \); }{\large \par}

{\large \( \widehat{\delta } \)\( \eta  \)=0;}{\large \par}

{\large \( \widehat{\delta } \)\( \overline{\eta } \)= - \( \frac{1}{g} \)\( \varepsilon ^{2}_{\eta } \){[}\( \partial  \).(B
+D)+ M\( (\varphi _{_{2}} \)'+\( \rho  \)\( _{2} \)){]}\( \delta \zeta  \);}{\large \par}

{\large \( \widehat{\delta } \)\( B_{\mu } \)=-\( \frac{1}{g} \)\( \varepsilon ^{2}_{B\mu \nu } \)\( \partial ^{\nu }(\eta +c) \)\( \delta \zeta  \);}{\large \par}

{\large \( \widehat{\delta } \)\( A_{\mu } \)=0,~~~~~~~~~~~~~~~~~~~~~~~~~~~(5.5a)}{\large \par}

{\large In addition, we shall find it convenient to employ the trivial symmetry
as in Ref. {[}6{]} .When it is added to those in (5.5a), the net symmetry transformations
read:}{\large \par}

{\large \( \widehat{\delta }\varphi _{1}' \)=2\( \varepsilon ^{2}_{\varphi _{1}'} \)\( (\varphi _{_{2}} \)'+\( \rho  \)\( _{2} \))(\( \eta +c \)
\( )\delta \zeta  \);}{\large \par}

{\large \( \widehat{\delta }\varphi _{2}' \)=\{-2\( \varepsilon ^{2}_{\varphi _{2}'} \)\( (\varphi _{_{1}} \)'+\( \rho  \)\( _{1} \))(\( \eta +c \)
\( )-2v\eta \}\delta \zeta  \); }{\large \par}

{\large \( \widehat{\delta } \)\( \eta  \)=0;}{\large \par}

{\large \( \widehat{\delta } \)\( \overline{\eta } \)= - \( \frac{1}{g} \){[}\( \partial  \).B+
M\( \varphi _{_{2}} \)'{]}\( \delta \zeta  \);}{\large \par}

{\large \( \widehat{\delta } \)\( B_{\mu } \)=-\( \frac{1}{g} \)\( \partial _{\mu }\eta  \)\( \delta \zeta  \);}{\large \par}

{\large \( \widehat{\delta } \)\( A_{\mu } \)=0,~~~~~~~~~~~~~~~~~~~~~~~~~~~(5.5b) }{\large \par}

{\large We note the essential similarity with the nonlocal Y-M theory:The} \emph{\large linear}
{\large parts of the net transformation are} \emph{\large same} {\large as the
local case, while the quadratic parts of the net transformation are unmodified
by the trivial transformation.We note that the measure for the nonlocal theory
will be determined} \emph{\large with respect to these non-local transformations
of (5.5b).}{\large \par}

{\large In an arbitrary gauge \( \xi  \), we follow the procedure of I {[}see
also ref.7{]} , and smear the fields with the quadratic forms of the} \emph{\large Feynman
gauge} {\large action of ( 5.1).The nonlocalized action then reads}{\large \par}

{\large \( \widehat{S} \)\( _{\xi } \)=\( \widehat{S} \)\( _{F} \)+\( \Delta  \)\( \widehat{S} \) }{\large \par}

{\large \( \Delta  \)\( \widehat{S} \) = \( \int  \)d\( ^{4}x \)\{-\( \frac{1}{2\xi } \)(\( \partial  \).\( \widehat{B} \)
+ \( \xi  \)M\( \widehat{\varphi _{2}'} \))\( ^{2} \)+\( \frac{1}{2} \)(\( \partial  \).\( \widehat{B} \)
+ M\( \widehat{\varphi _{2}'} \))\( ^{2} \)+ 2(1-\( \xi  \))gMv\( \int  \)d\( ^{4}x \)\( \widehat{\overline{\eta }} \)\( \widehat{\eta } \)}{\large \par}

{\large +2(1-\( \xi  \))gM\( \int  \)d\( ^{4}x \)\( (\varphi _{_{1}} \)'+\( \rho  \)\( _{1} \))(\( \overline{\eta }+\overline{c} \)
)(\( \eta +c \) \( ) \)\} ~~~~~~~~~~~~~~~~~~~~~~~~~~~(5.6)}{\large \par}

{\large We shall now establish the WT-identities for this action and discuss
the \( \xi  \)-independence of the physical observables in it. In order to
do this , we shall consider the generating functional}{\large \par}

{\large W{[}sources{]} = \( \int  \){[}D\( \varphi  \){]}exp \{ i\( \widehat{S_{\xi }} \)
+ {[}source terms{]}\} ~~~~~~~~~~(5.7)}{\large \par}

{\large where}{\large \par}

{\large {[}D\( \varphi  \){]}= {[}D\( \varphi '_{1} \){]}{[}D\( \varphi  \)\( '_{2} \){]}{[}D\( \psi  \){]}{[}D\( \overline{\psi } \){]}{[}D\( \eta  \){]}{[}D\( \overline{\eta } \){]}{[}D\( A_{\mu } \){]}{[}D\( B_{\mu } \){]}
~~~~~~~(5.7a)}{\large \par}

{\large and the source terms are}{\large \par}

{\large {[}source terms{]}=i\( \int  \)d\( ^{4}x \) {[}J\( ^{\mu } \)B\( _{\mu } \)+K\( ^{\mu } \)A\( _{\mu } \)+\( \chi ^{1}\varphi  \)\( '_{1}+\chi ^{2}\varphi  \)\( '_{2} \)+\( \overline{\lambda } \)\( \psi  \)+\( \overline{\psi } \)\( \lambda  \)+\( \overline{\theta } \)\( \eta  \)+\( \overline{\eta } \)\( \theta  \){]}
~~~~~~~~(5.7b)}{\large \par}

{\large We then have for the gauge variation of W with respect to the bare gauge
parameter \( \xi  \) keeping other} \emph{\large bare parameters fixed}\footnote{%
We shall account for the variation of these bare parameters with \( \xi  \)
later in sec.6
}\emph{\large ,}{\large \par}

{\large -i\( \frac{\partial W}{\partial \xi } \)=\( \int  \){[}D\( \varphi  \){]}exp
\{ i\( \widehat{S_{\xi }} \) + {[}source terms{]}\}\( \frac{\partial \Delta \widehat{S_{\xi }}}{\partial \xi } \)}{\large \par}

{\large =\( \int  \){[}D\( \varphi  \){]}exp \{ i\( \widehat{S_{\xi }} \)
+ {[}source terms{]}\}\{\( \frac{1}{2\xi ^{2}} \)\( \int  \)d\( ^{4}x \)\{(\( \partial  \).\( \widehat{B} \))\( ^{2} \)-\( \xi  \)\( ^{2} \)M\( ^{2}\widehat{\varphi _{2}'} \)\( ^{2} \))-2Mgv\( \int  \)d\( ^{4}x \)\( \widehat{\overline{\eta }} \)\( \widehat{\eta } \)
-2gM\( \int  \)d\( ^{4}x \)\( (\varphi _{_{1}} \)'+\( \rho  \)\( _{1} \))(\( \overline{\eta }+\overline{c} \)
)(\( \eta +c \) \( ) \) \}~~~~~~~~~~~~~~~~~~~~~~~~~~~(5.8)}{\large \par}

{\large We shall confine ourselves to the gauge variation around \( \xi  \)=1:}{\large \par}

{\large \( \frac{\partial W}{\partial \xi } \)\( \mid _{_{\xi =1}} \)=i\( \int  \){[}D\( \varphi  \){]}exp
\{ i\( \widehat{S_{\xi }} \) + {[}source terms{]}\}\{\( \frac{1}{2} \)\( \int  \)d\( ^{4}x \)\{(\( \partial  \).\( \widehat{B} \))\( ^{2} \)-M\( ^{2}\widehat{\varphi _{2}'} \)\( ^{2} \))-2Mgv\( \int  \)d\( ^{4}x \)\( \widehat{\overline{\eta }} \)\( \widehat{\eta } \)
-2gM\( \int  \)d\( ^{4}x \)\( (\varphi _{_{1}} \)'+\( \rho  \)\( _{1} \))(\( \overline{\eta }+\overline{c} \)
)(\( \eta +c \) \( ) \)\} ~~~(5.9)}{\large \par}

{\large =1/2\( \int  \)d\( ^{4}x \) \( \partial  \)\( _{\mu } \)\( \varepsilon  \)\( ^{-1}_{B} \)\( \frac{\delta }{\delta J_{\mu }(x)} \)\( \int  \){[}D\( \varphi  \){]}exp
\{ i\( \widehat{S_{F}} \) + {[}source terms{]}\}\( \varepsilon  \)\( ^{-1}_{B} \)\( \partial  \)\( .B \)(x)}{\large \par}

{\large -\( \frac{M^{2}}{2} \)\( \int  \)d\( ^{4}x \) \( \varepsilon  \)\( ^{-1}_{2} \)\( \frac{\delta }{\delta \chi _{2}(x)} \)\( \int  \)
{[}D\( \varphi  \){]}exp \{ i\( \widehat{S_{F}} \) + {[}source terms{]}\}\( \varepsilon  \)\( ^{-1}_{2} \)\( \varphi  \)\( '_{2} \)(x)}{\large \par}

{\large -2iMgv\( \int  \){[}D\( \varphi  \){]}exp \{ i\( \widehat{S_{\xi }} \)
+ {[}source terms{]}\}\( \int  \)d\( ^{4}x \)\{\( \widehat{\overline{\eta }} \)\( \widehat{\eta } \)+\( \frac{1}{v} \)\( (\varphi _{_{1}} \)'+\( \rho  \)\( _{1} \))(\( \overline{\eta }+\overline{c} \)
)(\( \eta +c \) \( ) \)\}~~~~~~(5.9a)}{\large \par}

{\large To simplify this further we use the WT identity}\footnote{%
As remarked earlier, our model has a chiral anomaly.It can be removed by adding
an additional fermion \( \Psi  \) of appropriate quantum numbers stated in
sec.3 .Then the WT below will contain additional terms that have to be carried
forward .They do not however alter the discussion of gauge-independence to one
loop qualitatively.
} {\large at \( \xi  \)=1:}{\large \par}

{\large 0 = \( \int  \){[}D\( \varphi  \){]}exp \{ i\( \widehat{S_{\xi }} \)
+ {[}source terms{]}\}\{\( \int  \)d\( ^{4}x \) {[} J\( ^{\mu } \)\( \widehat{\delta } \)B\( _{\mu } \)+K\( ^{\mu } \)\( \widehat{\delta } \)A\( _{\mu } \)+\( \overline{\lambda } \)\( \widehat{\delta } \)\( \psi  \)+\( \widehat{\delta } \)\( \overline{\psi } \)\( \lambda  \)+\( \chi ^{1}\widehat{\delta }\varphi '_{1}+\chi ^{2}\widehat{\delta }\varphi '_{2}+\overline{\theta }\widehat{\delta }\eta +\widehat{\delta }\overline{\eta }\theta  \){]}\( \mid _{\xi =1} \)~~~~~~~~~~~~~~~~(5.10)}{\large \par}

{\large To simplify the algebra, we shall set \( \overline{\lambda } \)=\( \lambda  \)=0;
as these terms do not play a nontrivial part in the algebraic manipulations
below.At the end, we shall put back the net contribution of these terms.This
gives }{\large \par}

{\large 0 = <\textcompwordmark{}<\( \int  \)d\( ^{4}x \) {[}- J\( ^{\mu } \)\( \partial _{\mu }\eta  \)+\( 2g\chi ^{1} \)\( \varepsilon ^{2}_{\varphi _{1}'} \)\( [(\varphi _{_{2}} \)'+\( \rho  \)\( _{2} \))(\( \eta +c \)
\( ) \){]}-2g\( \chi ^{2} \)\{\( \varepsilon ^{2}_{\varphi _{2}'} \)\( [(\varphi _{_{1}} \)'+\( \rho  \)\( _{1} \))(\( \eta +c \)
\( ) \)\( ]+v\eta \}+\theta  \){[}\( \partial  \).B+ M\( \varphi _{_{2}} \)'{]}
>\textcompwordmark{}>~~~~~~~~~~~~~~~(5.11)}{\large \par}

{\large where <\textcompwordmark{}< O >\textcompwordmark{}> stands for \( \int  \){[}D\( \varphi  \){]}O
exp \{ i\( \widehat{S_{\xi }} \) + {[}source terms{]}\}.Differentiating with
respect to \( \theta  \)(y) and setting \( \theta  \) = 0, we obtain the WT
identity useful for our purpose:}{\large \par}

{\large 0 = <\textcompwordmark{}<-i{[}\( \partial  \).B+ M\( \varphi _{_{2}} \)'{]}(y) }{\large \par}

{\large \( \qquad \qquad  \)+\( \int  \)d\( ^{4}x \){[}- J\( ^{\mu } \)\( \partial _{\mu }\eta  \)+2g\( \chi ^{1} \)\( \varepsilon ^{2}_{\varphi _{1}'} \){[}\( (\varphi _{_{2}} \)'+\( \rho  \)\( _{2} \))(\( \eta +c \)
\( ) \){]}-2g\( \chi ^{2} \)\{\( \varepsilon ^{2}_{\varphi _{2}'} \){[}\( (\varphi _{_{1}} \)'+\( \rho  \)\( _{1} \))(\( \eta +c \)
\( ) \)+ v\( \eta  \)\}{]}\( \overline{\eta (y)} \)>\textcompwordmark{}>
.~~~~~~~~(5.12)}{\large \par}

{\large We now operate on this by (-i/2) \( \partial ^{y}_{\mu } \)\( \frac{\delta }{\delta J_{\mu }(y)} \)\( \varepsilon _{\eta }^{-2} \)(y).}{\large \par}

{\large We find {[} note:\( \varepsilon _{B}^{-2} \)=\( \varepsilon _{\eta }^{-2} \)
{]}}{\large \par}

{\large 0 = <\textcompwordmark{}<-\( \frac{i}{2} \)\( \partial  \).B(y)\( \varepsilon _{B}^{-2} \){[}\( \partial  \).B+M\( \varphi _{_{2}} \)'{]}(y)
+\( \frac{1}{2} \) \( \int  \)d\( ^{4}x \){[}-J\( ^{\mu } \)\( \partial _{\mu }\eta  \)
+ {[}2g \( \chi ^{1} \)\( \varepsilon ^{2}_{\varphi _{1}'} \)\( (\varphi _{_{2}} \)'+\( \rho  \)\( _{2} \))(\( \eta +c \)
\( ) \)-2g\( \chi ^{2} \)\{\( \varepsilon ^{2}_{\varphi _{2}'} \){[}\( (\varphi _{_{1}} \)'+\( \rho  \)\( _{1} \))(\( \eta +c \)
\( ) \)+ v\( \eta  \)\}{]} \( \partial  \).B(y)\( \varepsilon _{\eta }^{-2} \)\( \overline{\eta } \)(y)
+\( \frac{i}{2} \)\( \partial  \)\( ^{2} \)\( \eta \varepsilon _{\eta }^{-2} \)\( \overline{\eta } \)(y)>\textcompwordmark{}>~~~~~~~~~(5.13)}{\large \par}

{\large Now operating by {[}\( \frac{Mi}{2} \) \( \frac{\delta }{\delta \chi _{2}} \)\( \varepsilon  \)\( _{\eta }^{-2} \){]}(y)
on (5.12),we obtain,}{\large \par}

{\large 0=<\textcompwordmark{}<\( \frac{iM}{2} \)\( \varphi _{_{2}} \)'\( \varepsilon ^{-2}_{B} \)
{[}\( \partial  \).B(y) + M\( \varphi _{_{2}} \)'(y){]}+ iMg\( \varepsilon _{\eta }^{-2} \)\( \overline{\eta } \)(y)\{\( \varepsilon ^{2}_{\varphi _{2}'} \){[}\( (\varphi _{_{1}} \)'+\( \rho  \)\( _{1} \))(\( \eta +c \)
\( ) \)(y){]}+v\( \eta  \) ) +\( \frac{M}{2} \)\( \varphi _{_{2}} \)'\( \varepsilon _{\eta }^{-2} \)\( \overline{\eta } \)(y)
\( \int  \)d\( ^{4}x \){[}-J\( ^{\mu } \)\( \partial _{\mu }\eta  \) + 2g\( \chi ^{1} \)\( \varepsilon ^{2}_{\varphi _{1}'} \)\( (\varphi _{_{2}} \)'+\( \rho  \)\( _{2} \))(\( \eta +c \)
\( ) \)-2g\( \chi ^{2} \)\{\( \varepsilon ^{2}_{\varphi _{2}'} \){[}\( (\varphi _{_{1}} \)'+\( \rho  \)\( _{1} \))(\( \eta +c \)
\( ) \){]}+ v\( \eta  \)\}{]} >\textcompwordmark{}> ..~~~~~~~~(5.14) }{\large \par}

{\large We now add (5.13) and (5.14) and integrate over d\( ^{4} \)y to obtain}\footnote{%
We recall \( \int  \) a \( \varepsilon  \)\( ^{-2} \)b =\( \int  \) b \( \varepsilon  \)\( ^{-2} \)a
=\( \int  \) \( \varepsilon  \)\( ^{-1} \)b \( \varepsilon  \)\( ^{-1} \)a
\( \quad  \)etc as easily seen in momentum representation.
} {\large ,}{\large \par}

{\large 0=<\textcompwordmark{}<\{-\( \frac{i}{2} \)\( \int  \)d\( ^{4} \)y\{(\( \partial  \).\( \widehat{B} \))\( ^{2} \)-M\( ^{2}\widehat{\varphi _{2}'} \)\( ^{2} \))(y)-\( \frac{i}{2} \)\( \widehat{\overline{\eta }} \)\( \partial ^{2} \)\( \widehat{\eta } \)+\( \frac{i}{2} \)M\( ^{2} \)\( \widehat{\overline{\eta }} \)\( \widehat{\eta } \)}{\large \par}
{\large +igM\( \varepsilon _{\eta }^{-2} \)\( \overline{\eta } \)(y)\( \varepsilon ^{2}_{\varphi _{2}'} \){[}\( (\varphi _{_{1}} \)'+\( \rho  \)\( _{1} \))(\( \eta +c \)
\( ) \){]} -\( \frac{1}{2} \)\( \varepsilon _{\eta }^{-2} \)\( \overline{\eta } \)(y){[}\( \partial  \).B(y)-M\( \varphi _{_{2}} \)'{]}(y)\( \int  \)d\( ^{4}x \){[}-J\( ^{\mu } \)\( \partial _{\mu }\eta  \)
+ 2g\( \chi ^{1} \)\( \varepsilon ^{2}_{\varphi _{1}'} \)\( (\varphi _{_{2}} \)'+\( \rho  \)\( _{2} \))(\( \eta +c \)
\( ) \)}{\large \par}

{\large -2g\( \chi ^{2} \)\{\( \varepsilon ^{2}_{\varphi _{2}'} \){[}\( (\varphi _{_{1}} \)'+\( \rho  \)\( _{1} \))(\( \eta +c \)
\( ) \){]}+ v\( \eta  \)\}{]}\}>\textcompwordmark{}> .~~~~~~~~(5.15)}{\large \par}

{\large We substitute (5.15) in (5.9) to obtain,}{\large \par}

{\large \( \frac{\partial W}{\partial \xi } \)\( \mid _{_{\xi =1}} \)=<\textcompwordmark{}<\{
-\( \frac{1}{2} \)\( \int  \)d\( ^{4} \)y \( \varepsilon _{\eta }^{-2} \)\( \overline{\eta } \)(y){[}\( \partial  \).B(y)-M\( \varphi _{_{2}} \)'{]}(y)\( \int  \)d\( ^{4}x \){[}-J\( ^{\mu } \)\( \partial _{\mu }\eta  \)
+ 2g\( \chi ^{1} \)\( \varepsilon ^{2}_{\varphi _{1}'} \)\( (\varphi _{_{2}} \)'+\( \rho  \)\( _{2} \))(\( \eta +c \)
\( ) \)-2g\( \chi ^{2} \)\{\( \varepsilon ^{2}_{\varphi _{2}'} \){[}\( (\varphi _{_{1}} \)'+\( \rho  \)\( _{1} \))(\( \eta +c \)
\( ) \){]}+ v\( \eta  \)\}{]}+ ghost terms>\textcompwordmark{}> ~~~~~~~(5.16)}{\large \par}

{\large where the} \emph{\large net} {\large ghost terms are }{\large \par}

{\large \( \int  \)d\( ^{4} \)y\{-\( \frac{i}{2} \)\( \widehat{\overline{\eta }} \)\( \partial ^{2} \)\( \widehat{\eta } \)+\( \frac{i}{2} \)M\( ^{2} \)\( \widehat{\overline{\eta }} \)\( \widehat{\eta } \)
+igM\( \varepsilon _{\eta }^{-2} \)\( \overline{\eta } \)(y) \( \varepsilon ^{2}_{\varphi _{2}'} \){[}\( (\varphi _{_{1}} \)'+\( \rho  \)\( _{1} \))(\( \eta +c \)
\( ) \){]}}{\large \par}

{\large -2igM{[}v\( \widehat{\overline{\eta }} \)\( \widehat{\eta } \) +\( (\varphi _{_{1}} \)'+\( \rho  \)\( _{1} \))(\( \overline{\eta }+\overline{c} \)
)(\( \eta +c \) \( ) \)(y){]} \} ~~~~~~~~(5.17) }{\large \par}

{\large These can be simplified using integration by parts in the third term
{[}\( \varepsilon _{\eta }^{2} \)=\( \varepsilon ^{2}_{\varphi _{2}'} \){]}.We
get}\footnote{%
We recall that while differentiating the \emph{net} \emph{action,}we need not
differentiate shadow fields in it as such terms vanish .
}{\large ,}{\large \par}

{\large \( \frac{i}{2} \)\( \overline{\eta } \)\( \frac{\delta \widehat{S_{F}}}{\delta \overline{\eta }} \)
- igM \( \overline{c} \) \( (\varphi _{_{1}} \)'+\( \rho  \)\( _{1} \))(\( \eta +c \)
\( ) \)~~~~~~~(5.18) }{\large \par}

{\large We shall now put back the fermion source terms that we had dropped earlier
in the intermediate algebra for our convenience.We thus arrive at the total
and simplified expression}\footnote{%
Recall the footnote earlier about the chiral anomaly.
} {\large for \( \frac{\partial W}{\partial \xi } \)\( \mid _{_{\xi =1}} \):}{\large \par}

{\large \( \frac{\partial W}{\partial \xi } \)\( \mid _{_{\xi =1}} \) =<\textcompwordmark{}<\{
-\( \frac{1}{2} \)\( \int  \)d\( ^{4} \)y \( \varepsilon _{\eta }^{-2} \)\( \overline{\eta } \)(y){[}\( \partial  \).B(y)-M\( \varphi _{_{2}} \)'{]}(y)}{\large \par}

{\large \( \bullet  \)\( \int  \)d\( ^{4}x \){[}-J\( ^{\mu } \)\( \partial _{\mu }\eta  \)
+ 2g\( \chi ^{1} \)\( \varepsilon ^{2}_{\varphi _{1}'} \)\( (\varphi _{_{2}} \)'+\( \rho  \)\( _{2} \))(\( \eta +c \)
\( ) \)-2g\( \chi ^{2} \)\{\( \varepsilon ^{2}_{\varphi _{2}'} \){[}\( (\varphi _{_{1}} \)'+\( \rho  \)\( _{1} \))(\( \eta +c \)
\( ) \){]}+ v\( \eta  \)\} + ig\( \overline{\lambda } \)\( \varepsilon
^{2}_{\psi } \) \( \gamma  \)\( _{5} \)(\( \psi  \)+\( \zeta  \))(\( \eta +c
\) \( ) \)+ ig\( \varepsilon ^{2}_{\psi } \)(\( \overline{\psi } \)+ \(
\overline{\zeta } \))(\( \eta +c \) \( ) \) \( \gamma  \)\( _{5} \)\( \lambda
\) {]}+ \( \frac{i}{2} \)\( \overline{\eta } \)\( \frac{\delta
\widehat{S_{F}}}{\delta \overline{\eta }} \) - igM \( \overline{c} \) \(
(\varphi _{_{1}} \)'+\( \rho  \)\( _{1} \))(\( \eta +c \) \( )
\)>\textcompwordmark{}> ~~~~~~~(5.19) }{\large \par} {\large We shall utilize
the above equation while discussing the \( \xi  \)-independence of the
S-matrix.}{\large \par}

\section{{\large GAUGE INDEPENDENCE OF THE S-MATRIX ELEMENTS}\large }

{\large In section 5 ,we have presented a modified way of constructing nonlocal
field theory for an arbitrary \( \xi  \), for the spontaneously broken model,
by analogy with our work in I. We expect from our results in I, that unlike
the usual scheme of Ref.6, the S-matrix elements with this modified regularization
are now \( \xi  \)-independent.In the last section, we formulated the WT identity
for the model and obtained an expression for \( \frac{\partial W}{\partial \xi } \)\( \mid _{_{\xi =1}} \)
in a simplified form.We shall now use it to verify the gauge-independence in
a limited sense: in one-loop approximation and around \( \xi  \)=1}\footnote{%
We recall that the possible gauge-dependent part of S-matrix elements is intrinsically
at least one loop; so we can regard within it \( \xi  \)\( _{R} \)= \( \xi  \)\( _{UR} \)=1
to this order. So we need not distinguish between the two.
}{\large . We shall find it necessary to construct a way to deal with the extra
ghost term in (5.19), that, as we shall see, is specific to the nonlocal theories.
We shall do this in one-loop approximation.}{\large \par}

{\large To this end, we shall first make a number of observations: }{\large \par}

{\large (a) We have evaluated \( \frac{\partial W}{\partial \xi } \) keeping
the bare quantities {[}}\emph{\large bare} {\large sources,}\emph{\large bare}
{\large masses and couplings{]} fixed. To show gauge-independence, we need to
first evaluate \( \frac{\partial W}{\partial \xi } \)\( \mid _{_{R}} \) i.e.
the derivative with} \emph{\large renormalized} {\large parameters} \emph{\large }{\large and
sources held fixed. Denoting by S\( ^{R} \) and p\( ^{R} \) the renormalized
sources and parameters {[}masses and couplings{]}, and by S\( ^{UR} \) and
p\( ^{UR} \), the unrenormalized ones,we can write: }{\large \par}

{\large \( \frac{\partial W}{\partial \xi } \)\( \mid _{_{R}} \) = \( \frac{\partial W}{\partial \xi } \)\( \mid _{_{UR}} \)
+ \( \sum  \)\( \frac{\partial S^{UR}}{\partial \xi } \)\( \mid _{_{R}} \)
\( \frac{\delta W}{\delta S^{UR}} \) +\( \sum  \)\( \frac{\partial p^{UR}}{\partial \xi } \)\( \mid _{_{R}} \)
\( \frac{\delta W}{\delta p^{UR}} \)\( \qquad \qquad \qquad \qquad \qquad  \)(6.1)}{\large \par}

{\large To show gauge independence, we have to show that the contribution to
an S-matrix element from \( \frac{\partial W}{\partial \xi } \)\( \mid _{_{UR}} \)
is canceled by the last two terms.We verify this at \( \xi  \)=1. For this
purpose, we make several observations on (5.19). }{\large \par}

{\large (i) We can ignore the \( \partial  \).J term as it does not contribute
to Green's functions with physical polarization vectors {[}\( \epsilon  \).p
=0{]} attached. }{\large \par}

{\large (ii) We can set \( \chi  \)\( _{2} \) =0, as it refers to an unphysical
scalar.We can also set sources for the ghost fields to zero for similar reasons.}{\large \par}

{\large (iii) We can set <\textcompwordmark{}<\( \overline{\eta } \)\( \frac{\delta \widehat{S_{F}}}{\delta \overline{\eta }} \)>\textcompwordmark{}>
to zero by the ghost equation of motion.There, we need to recall that the one-loop
measure is independent of \( \overline{\eta } \) and the Jacobian term is a
constant independent of sources and does not contribute to Green's functions
{[}See e.g. appendix of Ref. 9{]}. }{\large \par}

{\large (iv) The source terms proportional to \( \chi  \)\( _{1} \),\( \lambda  \)
and \( \overline{\lambda } \) are of the same kind as they appear in the discussion
of the gauge-independence of the local theory {[}12{]}; except that they have
a modified appearance on account of non-locality.The treatment of these terms
is very similar to that for the local theory. To sum up these terms contribute
to (a) the two-point functions of the respective fields and this contribution
is related to the \( \xi  \)-dependence of the 2-point function of the wave-function
of the renormalization;(b) to the on-shell physical Green's functions in
the form of} \emph{\large external line insertions only} {\large ( barring {}``exceptional
momenta{}'' {[}13{]}) (c) the net contribution to an S-matrix element is canceled
by the second term on the right hand side of (6.1) which has the \( \xi  \)-dependence
of Z's. }{\large \par}

{\large That leaves us with a term of the term:}{\large \par}

{\large - 2igM\( \int  \)d\( ^{4} \)y \( \overline{c} \) \( (\varphi _{_{1}} \)'+\( \rho  \)\( _{1} \))(\( \eta +c \)
\( ) \) \( \qquad \qquad \qquad \qquad \qquad  \)(6.2)}{\large \par}

{\large We note that if the local limit (\( \Lambda  \)\( \rightarrow \infty  \))
is taken first, this term would vanish {[} see (6.4) below{]}; and thus is a
new type of term that arises only in the nonlocal model.We shall show that this
term does contribute to the 2-point function of the scalar \( \varphi _{_{1}} \)',
and corresponds to a (finite) \( \xi  \)-dependence of the bare mass. In higher
point on-shell physical Green's function, it contributes again as ghost tadpole
insertions and is similarly canceled by opposite contributions from the last
term in (6.1).}{\large \par}

{\large By power counting,the above term is an operator of dimension 3. As we
shall see, moreover, it can be cast as operator of an effective dimension 2.
To see this,we recall:}{\large \par}

{\large \( \overline{c} \)={[}\( \frac{\varepsilon _{\eta }^{2}-1}{\Im _{\eta }} \){]}\( \frac{\delta ^{P}I[\phi +\psi ]}{\delta \eta } \)\( \mid  \)
\( \qquad \qquad \qquad \qquad \qquad  \)(6.3)}{\large \par}

{\large where \( \delta  \)\( ^{P} \) denotes partial derivative with respect
to a {}``\( \phi  \){}'' keeping {}``\( \psi  \){}'' fixed; and the vertical
line implies that the result is to be evaluated at \( \psi  \)=\( \psi  \){[}\( \phi  \){]}.
We can simplify this as }{\large \par}

{\large \( \overline{c} \) = 2 Mg {[}\( \frac{\varepsilon _{\eta }^{2}-1}{\Im _{\eta }} \){]}\{\( (\varphi _{_{1}} \)'+\( \rho  \)\( _{1} \))(\( \overline{\eta }+\overline{c} \)
)\} \( \qquad \qquad \qquad \qquad \qquad  \)(6.4)}{\large \par}

{\large {[} as noted earlier,\( \overline{c} \)\( \rightarrow  \)0 as \( \Lambda  \)\( \rightarrow \infty  \){]}.Thus,
this extra term in (6.2) becomes, }{\large \par}

{\large i(2Mg)\( ^{2} \)\( \int  \)d\( ^{4} \)y {[}\( \frac{\varepsilon _{\eta }^{2}-1}{\Im _{\eta }} \){]}\{\( (\varphi _{_{1}} \)'+\( \rho  \)\( _{1} \))(\( \overline{\eta }+\overline{c} \)
)\}\( (\varphi _{_{1}} \)'+\( \rho  \)\( _{1} \))(\( \eta +c \) \( ) \)\( \qquad \qquad \qquad \qquad \qquad  \)(6.5) }{\large \par}

{\large where we note} \emph{\large two} {\large powers of M before the {[}nonlocal{]}
operator. }{\large \par}

{\large We note several things about the operator in (6.5).}{\large \par}

{\large (i) We need to consider the Green's functions only with physical external
lines.Then the ghost fields above must necessarily be contracted (in a loop).
Then the factor {[}\( \frac{\varepsilon _{\eta }^{2}-1}{\Im _{\eta }} \){]},
in momentum space, involves an} \emph{\large internal} {\large
momentum, and thus counts for (-2) momentum dimensions in such diagrams.Thus,
it behaves, for power counting purposes, an operator of dimension}
\emph{\large two}{\large . }{\large \par}
{\large (ii) The only two-point function of physical fields,to which it can
contribute in one loop is the two-point function of \( \varphi _{_{1}} \)'.The
corresponding diagram, is by naive power counting, log-divergent. However ,
in view of the fact that {[}\( \frac{\varepsilon _{\eta }^{2}(p)-1}{\Im _{\eta }(p)} \){]}\( \rightarrow  \)
0 as \( \Lambda  \)\( ^{2} \)\( \rightarrow  \) \( \infty  \), it is rendered
convergent, as a direct calculation also confirms.It leads to a} \emph{\large finite}
{\large contribution to the \( \xi  \)-dependence of the mass of \( \varphi _{_{1}} \)'. }{\large \par}

{\large (iii) When we consider the insertion of this operator in higher point
Green's functions in one loop approximation,we note that there are two kinds
of diagrams: one in which the ghost loop has no additional vertices and the
other in which the ghost loop has additional lines emerging out of it. The second
class of diagrams contain {[}1-loop{]} 1PI subgraphs with more than 2 external
lines.These subgraphs are convergent by naive power counting itself. In these
diagrams, the limit \( \Lambda  \)\( ^{2} \)\( \rightarrow  \) \( \infty  \)
can be taken inside.But then as pointed out the diagram contains a factor {[}\( \frac{\varepsilon _{\eta }^{2}(p)-1}{\Im _{\eta }(p)} \){]}
that vanishes as \( \Lambda  \)\( ^{2} \)\( \rightarrow  \) \( \infty  \).
Thus , such proper subgraphs {[}and therefore such one loop diagrams which have
no other divergences{]} vanish as \( \Lambda  \)\( ^{2} \)\( \rightarrow  \)
\( \infty  \) is taken in them. {[}This has been explicitly verified for the
3-\( \varphi _{_{1}} \)' and the 4-\( \varphi _{_{1}} \)' proper
vertices{]}. Also the diagrams having a ghost tadpole ,a finite quantity,
attached to a vertex {[}rather than a line{]}, which can arise when a
shadow field in the operator is expended, also vanish as \( \Lambda  \)\( ^{2}
\)\( \rightarrow  \) \( \infty  \). Therefore, the only diagrams for higher
point functions, in which this operator insertion matters are those with a
ghost tadpole insertion on any one \( \varphi _{_{1}} \)' line {[}internal or
external{]}. Such contributions to \( \frac{\partial W}{\partial \xi } \)\(
\mid _{_{UR}} \) are canceled by opposite contributions to \( \sum  \)\(
\frac{\partial p^{UR}}{\partial \xi } \)\( \mid _{_{R}} \) \( \frac{\delta
W}{\delta p^{UR}} \) with the parameter '\( p \) ' referring to the bare mass
of \( \varphi _{_{1}} \)'. We further note that in view of the fact that there
are no higher point 1PI graphs surviving as \( \Lambda  \)\( ^{2} \)\(
\rightarrow  \) \( \infty  \), correspondingly, there is no \( \xi
\)-dependence induced by this term in bare couplings and \( \sum  \)\(
\frac{\partial p^{UR}}{\partial \xi } \)\( \mid _{_{R}} \) \( \frac{\delta
W}{\delta p^{UR}} \) does not have nonvanishing terms when ' \( p \) ' refers
to any other parameter.}{\large \par} \section{CONCLUSIONS} {\large We finally
summarize our results. We considered, in this work , the elegant nonlocal
formulations of gauge theories of Ref.6 and applied it to a simple abelian
model with spontaneously broken symmetry and in R\( _{\xi } \) -gauges. We
applied the process of constructing the nonlocal theory with a finite \(
\Lambda  \) starting from the R\( _{\xi } \) -gauge local Lagrangian. We chose
a theory with both the vector and axial couplings, as this mimics the Standard
Model and conveys our results in a simpler context. We found by explicit
calculation that when the formulation of nonlocalization of Ref.{[}6{]} was
applied to such a model, it lead to a \( \xi  \)-dependent physical
observable, the (g-2) for the fermion. {[} A similar result had also been
checked {[}10{]} in the Standard Model earlier{]}. We then applied the
alternate procedure of nonlocalization suggested and tested for the unbroken
nonabelian gauge theories {[}9{]}.We formulated the WT identities for this
model. As a test of the modified regularization scheme, we also verified the
\( \xi \)-independence of observable S-matrix elements in one loop order in
the vicinity of the Feynman gauge.}{\large \par}  
\begin{thebibliography}{10}
\bibitem{1}{\large See for example, Ta-Pei Cheng and Ling-Fong Li in {}``Gauge
Theories of Fundamental Interactions{}'' (Oxford:Clarendon) 1984}{\large \par}
\bibitem{2}{\large G. 't Hooft and M. Veltman Nucl. Phys. B33, 189
(1972)}{\large \par} \bibitem{3}{\large J.Moffat Phys. Rev.
D41,1177(1990)}{\large \par} \bibitem{4}{\large E. D. Evans et al, Phys Rev
D43, 499 (1991)}{\large \par} \bibitem{5}{\large G. Kleppe and R. P. Woodard,
Ann. Phys. (N.Y.) 221, 106 (1993)}{\large \par} \bibitem{6}{\large G. Kleppe
and R. P. Woodard, Nucl. Phys. B388, 81 (1992)}{\large \par}
\bibitem{7}{\large M. A. Clayton Gauge Invariance in nonlocal regularized QED.
(Toronto U.) UTPT-93-14, Jul 1993. {[}hep-th/9307089{]}}{\large \par}
\bibitem{8}{\large See e.g.N. J. Cornish Mod.Phys.Letts.A 7,1895 (1992);
M.A.Clayton et al Int. Jour. Mod. Phys.A 9, 4549 (1994) and references
therein.}{\large \par} \bibitem{9}{\large Anirban Basu and S.D.Joglekar
J.Math.Phys. 41,7206 (2000) }{\large \par} \bibitem{10}{\large S. D. Joglekar
and G. Saini,1996 (Unpublished)}{\large \par} \bibitem{11}{\large S. D.
Joglekar and G. Saini, Z. Phys. C.76, 343-353 (1997)}{\large \par}
\bibitem{12}{\large See e.g. B. W. Lee in {}``Methods in field theory{}''Les
Houches 1975; Editor R. Balian and J.Zinn-Justin}{\large \par}
\bibitem{13}{\large For a detailed explanation regarding exceptional momenta
and why we can ignore gauge dependence for them ,see e.g. S.D.Joglekar
Ann.Phys. 109,210(1977).}{\large \par} \bibitem{14}{\large S.D.Joglekar
hep-th/0003104 J.Phys.A (2001) to appear;and S.D.Joglekar hep-th/0003077
}{\large \par}\end{thebibliography}
\end{document}